\documentclass[]{aa}
\usepackage{times}
\usepackage{natbib}
\usepackage[dvips]{graphicx}
\usepackage{epsfig,longtable,lscape}
\usepackage{epsfig}
\usepackage[figuresright]{rotating}
\bibpunct{(}{)}{;}{a}{}{,}

\usepackage{float}
\usepackage{amssymb}
\usepackage{amsmath}
\usepackage{aas_macros}
\usepackage{headings}
\usepackage{latexsym}
\usepackage{graphics}
\usepackage{graphicx}

\def\1E{1E1207.4$-$5209}
\def\XMM{{\em XMM--Newton}}
\def\EPICn{{\em European Photon Imaging Camera}}
\def\EPIC{{\em EPIC}}
\def\MOSn{{\em Metal Oxide Semi--conductor}}
\def\MOS{{\em MOS}}

\def\pn{{\em pn}}
\def\WFIn{{\em Wide Field Imager}}
\def\WFI{{\em WFI}}

\def\SAS{{\em SAS}}

\def\gsc{{\em GSC}}
\def\tmass{{\em 2MASS}}

\def\theli{{\em THELI}}

\def\ltsima{$\; \buildrel < \over \sim \;$}
\def\simlt{\lower.5ex\hbox{\ltsima}}
\def\gtsima{$\; \buildrel > \over \sim \;$}
\def\simgt{\lower.5ex\hbox{\gtsima}}

\begin{document}

\title{A Deep \XMM\ Serendipitous Survey of a middle--latitude area. \\ II. New deeper X-ray and optical observations.\thanks{Based on observations collected at ESO, La Silla, under Programmes 073.D-0621(A) and 074.D-0613(A)}}

\author{G. Novara\inst{1,2}, N. La Palombara\inst{1}, R. P. Mignani\inst{3}, E. Hatziminaoglou\inst{4}, M.Schirmer\inst{5,6}, A. De Luca\inst{1,2,7}, P.A. Caraveo\inst{1,2}}

\institute{INAF-IASF, Istituto di Astrofisica Spaziale e Fisica Cosmica ``G.Occhialini'', Via Bassini 15, I--20133, Milano, Italy
 \and Universit\`a di Pavia, Dipartimento di Fisica Teorica e Nucleare, Via Ugo Bassi 6, I--27100, Pavia, Italy
 \and Mullard Space Science Laboratory, University College London, Holmbury St Mary, Dorking, RH56NT, Dorking, UK 
 \and European Southern Observatory, Karl Schwarzschild Str.2, D--85748, Garching, Germany 
 \and Isaac Newton Group of Telescopes, Apartado de correos 321, S--38700, Santa Cruz de La Palma Tenerife, Spain 
 \and Argelander-Institut f\"ur Astronomie, Auf dem H\"ugel 71,  D--53121, Bonn, Germany
 \and IUSS - Istituto Universitario di Studi Superiori, Viale Lungo Ticino Sforza 56, I--27100, Pavia, Italy.      }

\titlerunning{\XMM\ observations of a middle-latitude field}
\authorrunning{Novara et al.}

\offprints{Giovanni Novara, novara@iasf-milano.inaf.it}

\date{Received 16 July 2008 / Accepted 21 March 2009}

\authorrunning{G. Novara et al.}

\titlerunning{A Deep \XMM\ Serendipitous Survey}

\abstract{The radio--quiet neutron  star 1E1207.4$-$5209 has been the target  of several \XMM\ observations, with a total exposure of $\sim$ 350 ks. The source is located at intermediate galactic  latitude ($b \sim 10^{\circ}$), i.e. in a sky region with an  extremely interesting mix of both galactic and extra-galactic X--ray sources.}  {The aim of our
work    is   to    investigate    the   properties    of   both    the
intermediate-latitude   galactic  and  extra-galactic   X--ray  source
populations in the \1E\ field.}   {We performed a coherent analysis of
the  whole  \XMM\  observation  data  set  to  build  a  catalogue  of
serendipitous  X--ray sources  detected  with high  confidence and  to
derive information on the  source flux, spectra, and time variability.
In addition, we performed  a complete multi-band ({\em UBVRI}) optical
coverage  of the  field with  the \WFIn\  (\WFI) of  the  ESO/MPG 2.2m
telescope (La  Silla) to search for candidate  optical counterparts to
the  X--ray sources,  down to  a V-band  limiting magnitude  of $\sim$
24.5.}  {From the combined observation data set we detected a total of
144  serendipitous X--ray  sources.  We find evidence that  the source
log$N$--log$S$  distribution may  be  different from  those
computed either in  the Galactic plane or at  high galactic latitudes.
Thanks to the refined X--ray  positions and to the \WFI\ observations,
we found candidate optical counterparts for most of the X--ray sources
in  our compilation.  For  most of  the brightest  ones we  proposed a
likely classification based on both the X--ray spectra and the optical
colours.}  {Our results indicate that at  intermediate galactic
latitude   the  X--ray   source   population  is   dominated  by   the
extra--galactic  component, but with  a significant  contribution from 
the galactic component in the soft energy band, below 2 keV.}


\maketitle

\section{Introduction}

Since the launch  of \XMM\ in 1999, the radio--quiet neutron star \1E\
in the  supernova remnant (SNR) PKS  1209$-$51 has been  the target of
several  observations,  for  a   total of  $\sim$  450  ks  scheduled
time. Therefore, observations of this field make up one of the deepest
pencil-beam X--ray  surveys obtained  at  intermediate galactic
latitude ($|b|\simeq  10^{\circ}$). This gives  the unique opportunity
to sample, in  the same survey, both the  galactic and extra--galactic
X--ray  source population.   Thanks  to the  wide  energy range,  high
throughput,  and  good  spectral  resolution of  the  \EPICn\  (\EPIC)
\citep{Turner2001}, this  data set allows us to  investigate with high
sensitivity  both  the  distant  population of  quasi-stellar  objects
(QSOs), active  galactic nuclei (AGNs), normal galaxies,
and the galactic population of stars and X--ray binaries (XRBs).

The  two longest  \XMM\  observations, performed  in  August 2002  and
corresponding to  a total  of $\sim$ 260  ks of net  integration time,
were used to  study the pulsations and the  absorption features of the
neutron star  \1E\ \citep{Bignami2003,DeLuca2004}.  As  a by--product,
we  used the  data  of  the two  \MOSn\  (\MOS)  cameras to
investigate the  population of  the faint {\em  serendipitous} sources
detected  in  the  field.   This  yielded the  detection  of  196
serendipitous X--ray sources  (\citet{Novara2006}, hereafter Paper I),
which  were   characterised  by  a   very  interesting  log$N$--log$S$
distribution.  On the one hand, in the 0.5--2 keV energy range it
shows  an excess  with  respect to  both  the Galactic  plane and  the
high--latitude  distributions,  which   suggests  a  mixed  population
composed of  both galactic and extra--galactic sources.   On the other
hand, in the 2--10 keV  energy band the log$N$--log$S$ distribution is
comparable to that derived at high galactic latitudes, thus suggesting
that it is dominated  by extra--galactic sources.  The cross--match of
the list of serendipitous X--ray sources with version 2.3 of
the  \textit{Guide  Star  Catalogue}  (\textit{GSC  2.3})  \citep{Lasker2008} 
provided  a candidate optical counterpart for  about half of
them, down  to limiting magnitudes  $B_{J}\sim$ 22.5 and  $F\sim$ 20.
For  the 24 brightest  sources it  was possible  to obtain  a spectral
characterisation,  and  an  optical  identification was  proposed  for
$\sim$80\% of  them. Finally,  the detailed spectral  investigation of
one  of the  brightest  sources, characterised  by  a highly  absorbed
spectrum   and  an  evident   Fe  emission   line,  and   its  optical
identification with  the galaxy ESO 217-G29, led to it
being classified as a new Seyfert--2 galaxy.

These results prompted  us to extend our analysis  to the whole sample
of  the \XMM\  observations of  the \1E\  field.  In  addition  to the
observations    published    in    \citet{Bignami2003} and
\citet{DeLuca2004}, we  thus considered also the  first observation of
the field, performed in  December 2001 \citep{Mereghetti2002}, and the
sequence of the  seven observations, performed during a  40 day window
between June and  July 2005 \citep{Woods2007}.  In this  way we almost
doubled  the total  integration time  and significantly  increased the
count statistics. We  used this enlarged data set  to refine the study
of the serendipitous  X--ray source population.  We also
took  advantage  of the  improvements  of  the  \XMM\ data  processing
pipeline, which  now minimises the  number of spurious  detections and
provides  improved source position errors \citep{Watson2009}. Moreover,  we  performed
dedicated follow-up optical observations with the \WFIn\ (\WFI) of the
ESO/MPG 2.2m telescope down to $V \simeq 24.5$, i.e.  with a factor of
10 improvement in flux limit  compared to the \textit{GSC 2.3} used in
Paper I.

The paper  is organised as  follows: the X--ray observations  and data
reduction  are  described in  \S~\ref{sec:2},  while the  serendipitous
source  catalogue  and  the  analysis  of its  bright  subsample  are
presented   and   discussed   in  \S~\ref{sec:3}   and   \S~\ref{sec:4},
respectively. The optical observations and data analysis are described
in \S~\ref{sec:5} and the cross--correlations of the X--ray and optical
catalogues   is  described   in   \S~\ref{sec:6}.  The   optical/X--ray
classification of  the brightest sources,  as well as of  the peculiar
Seyfert--2 galaxy, are discussed in \S~\ref{sec:7}.

\section{X-Ray observations and data processing}\label{sec:2}

\subsection{Observations}

\1E\  was observed  with \XMM\  in ten  different pointings  from 2001
December 23 to 2005 July 31, for a net exposure time of $\sim$ 346 ks.
All      the      three      \EPIC\      focal      plane      cameras
\citep{Turner2001,Struder2001} were active during these pointings: the
two \MOS\ cameras were operated  in standard {\em Full Frame} mode, in
order to  cover the whole 30\arcmin\ field--of--view;  the \pn\ camera
was operated in {\em Small Window} mode, where only the on--target CCD
is read--out, in order to time--tag the individual photons and provide
accurate  arrival  time  information.   Since  we  are  interested  in
serendipitous  X--ray  sources  only,  in the  following  analyses  we
consider only data taken with the \MOS\ cameras.

In  Table~\ref{tab:expo} we  report the  \textit{good  time intervals}
(\textit{GTI})  of  the  two  \MOS\   cameras  for  each  of  the  ten
observations, i.e. the ``effective'' exposure times computed after the
soft--proton  rejection  (see  next  subsection). 
In  the  seven  2005 observations  the CCD  number 6  of the  \textit{MOS1} 
camera  was not active, since it was switched off in March 2005 due to a micrometeorite impact
\footnote{http://xmm.vilspa.esa.es/external/xmm\_news/items/MOS1-CCD6/index.shtml}.
For the second and the third observations of Table~\ref{tab:expo} both
\MOS\ cameras were used with  the thin filter, while the medium filter
was used for all the other observations.

\subsection{Data processing} 

For each pointing we obtained two data sets (i.e. one for each
\MOS\ camera), which  we processed independently through the
standard \XMM\  {\em Science  Analysis Software} ({\em  SAS}) v.7.1.0.
In the first step, the \XMM\  {\em SAS} tasks {\tt emproc} was used to
linearize the \MOS\ event files.  In the second step, event files were
cleaned up  for the effects of  soft protons flares.   We filtered out
time  intervals  affected by  high  instrument  background induced  by
flares of  soft protons  (with energies less  than a few  hundred keV)
hitting  the detector surface.  In order  to avoid  contributions from
genuine X--ray source variability, we selected only single and double
events (PATTERN$\leq$4) with energies  greater than 10 keV and
recorded  in   the  peripheral  CCDs  (CCD=2-7).   Then,   we  set  a
count--rate  threshold  for good  time  intervals  (GTI)  at 0.22  cts
s$^{-1}$. By selecting only events within GTIs we finally obtained two
``clean''  event lists for  each \MOS\  data set,  whose ``effective''
exposure times are reported in Table~\ref{tab:expo}.

\begin{table*}
\begin{center}
\caption[]{Log of the \XMM\ observations of the \1E\ field with the corresponding net 
        good time interval (GTI) for the two \EPIC/\MOS\ cameras.} 
\label{tab:expo}
\footnotesize{
\begin{tabular}{ccccc} \hline\noalign{\smallskip} \hline\noalign{\smallskip}
Observation ID & \XMM & Date & \multicolumn{2}{c}{GTI (ks)} \cr
       & revolution & (UT) & MOS1 & MOS2 \cr
\noalign{\smallskip\hrule\smallskip}
 0113050501 & 374 & 2001-12-23T18:59:41 & 24.3 & 25.2   \cr
 0155960301 & 486 & 2002-08-04T07:25:09 & 105.3 & 105.8 \cr
 0155960501 & 487 & 2002-08-06T07:17:29 & 100.7 & 102.0 \cr
\hline
 0304531501 & 1014 & 2005-06-22T12:10:05 & 15.1 & 15.1   \cr
 0304531601 & 1020 & 2005-07-05T00:44:58 & 18.3 & 17.9   \cr
 0304531701 & 1023 & 2005-07-10T06:43:47 & 7.1 & 9.3     \cr
 0304531801 & 1023 & 2005-07-11T02:00:45 & 56.6 & 54.5   \cr
 0304531901 & 1024 & 2005-07-12T11:08:22 & 3.5 & 3.2     \cr
 0304532001 & 1026 & 2005-07-17T00:18:21 & 12.7 & 10.7   \cr
 0304532101 & 1033 & 2005-07-31T14:03:09 & 2.5 & 2.1     \cr
\hline\noalign{\smallskip} \hline\noalign{\smallskip}
\end{tabular}}
\end{center}
\end{table*}

\subsection{Source detection}\label{SourceDet}

The \EPIC\ images of the \1E\ field show the presence of several faint
X--ray  sources. Therefore,  we used a source  detection  algorithm in
order to  produce a catalogue  of the serendipitous X--ray  sources in
the field. 

We decided to perform  the source detection in three different
energy bands:  the two standard  coarse soft/hard energy  bands 0.5--2
keV and 2--10 keV, and the total energy band 0.3--8 keV. First of all,
for each observation in Table~\ref{tab:expo} we used the cleaned event
file to produce {\em MOS1} and {\em MOS2} images in the three selected
energy bands together with the associated exposure maps, and hence
accounted for variations in spatial quantum efficiency {\em (QE)}, mirror
vignetting and effective field of view.

Although it would be interesting to look for variability on short time
scales,  we did  not run  the  source detection  for each  of the  ten
observations   in  Table~\ref{tab:expo}.    Indeed,   with  the 
exceptions  of   the  2002  observations   and the   fourth  2005
observation, all observations have too short an integration time 
to allow  for a  statistically significant time  variability analysis.
As  seen   from  Table~\ref{tab:expo},   we  thus  divided   the  full
observation  set in  two time  windows: the  first spanning  from 2001
December 23 to 2002 August 6 (three observations), the second spanning
from 2005 June  22 to 2005 July 31 (seven  observations).  We then ran
the  source  detection  on  each  of  these  two  observation  subsets
separately,  in  order to  search  for  long  term source  variability
(\S~\ref{sec:variab33}).

We merged the cleaned event files of the 2001/2002 and 2005
observation  subsets separately  to obtain,  for each  of  them, three
co--added images  in the three selected energy  bands. Since different
observations correspond  to different pointings,  which have different
aspect solutions,  we corrected, for  each of the three  energy bands,
the coordinates measured on the single observation {\em MOS1} and {\em
MOS2} exposure  maps   through  a  relative coordinate  transformation.  
To  this aim,  for each  of the  two time
windows we  selected the image with  the longest exposure  time and we
took it as a reference to register  all {\em MOS1} and {\em MOS2} exposure maps. 
We used the {\em IRAF} task
{\tt wregister} to compute the coordinate transformation and apply the
frame registration. In this way, for each energy band we merged the exposure
maps of each observation and \MOS~camera, thus obtaining total  exposure  maps  
corresponding to the co-added images built from the merged event file. 
For each observation subset, we then used three co-added images, one for each 
defined energy band,  and the corresponding total exposure maps as 
input to run the source detection.
Finally, we applied the same procedure to combine all the ten observations
Table~\ref{tab:expo}, so as to maximize the {\em signal--to--noise
(S/N)} ratio.
Below, we  give    details
about the procedure used  to run  the source  detection, for
each  energy  band,  in each  of  the  three  final data  sets:  those
corresponding   to   the   2001/2002   and   the   2005   observations
(Table~\ref{tab:expo}) and that  corresponding to the full observation
set.

\begin{enumerate}
\item For  each data set,  and for each  energy band, we run  the {\em
SAS} task {\tt eboxdetect} in {\em local mode} to create a preliminary
source list.   Sources were identified  by applying the  standard {\em
minimum  detection  likelihood}  criterion,  i.e.  we  validated  only
candidate  sources with  detection  likelihood {\em  -ln~P} $\ge$  8.5
\citep{Novara2006},  where {\em P}  is the  probability of  a spurious
detection due  to a Poissonian  random fluctuation of  the background.
This  corresponds to  a probability  $P=$ 2$\times$10$^{-4}$  that the
source count number in a given energy band originates from a background
fluctuation.   This implies  a  contamination of  at  most 1  spurious
source per energy band.

\item  Then, the  task  {\tt esplinemap}  was  run to  remove all  the
validated sources from  the original image and to  create a background
map by fitting the so called {\em cheesed image} with a cubic spline.

\item  For each  data set,  and for  each energy  band, the  task {\tt
eboxdetect} was run again in {\em  map mode}, using as a reference the
computed  background map.  For  each set,  the likelihood  values from
each individual  energy band were  then added and transformed  to {\em
equivalent single  band} detection likelihoods, and  a threshold value
of 8.5 was applied to accept or reject a detected source.
\end{enumerate}

Unfortunately, even using the maximum number of spline nodes (20), the
fit performed  in step 2 (see  above) is not  sufficiently flexible to
model the  local variations of the  background, due to  the presence of
the bright  SNR PKS  1209$-$51.  Therefore,  it was necessary  to correct
each background map {\em pixel by pixel}, measuring the counts both in
the {\em cheesed  image} and in the background  map itself by applying
the correction  algorithm described in  \citet{Baldi2002}. All sources
were then checked against the  corrected background maps and all their
parameters  calculated  again.  Finally,  for  each  energy band,  the
revised source list was filtered  to include, again, only sources with
corrected detection likelihood {\em -ln~P} $>$ 8.5.

\subsection{Source list}\label{SourceList}

At the end of the source  detection process we thus produced, for each
of the  three observation sets,  a master list including  only sources
with detection likelihood{\em  -ln~P} $>$8.5 in {\em at  least} one of
the three energy bands and  manually screened to reject residual false
detections.   For  each  source,  the  master  list  provides  various
parameters including the  detector and sky coordinates, the effective
exposure   time   and,   for   each   of  the   three   energy   bands
(soft/hard/total), the total counts,  count--rate and errors, the {\em
S/N} ratio,  and the detection  likelihood.  The master list  does not
include quantitative information on the source extension, which can be
used  for a preliminary  morphological classification  (point--like or
extended). This  is because the  significant distortion of the  PSF at
large   off--axis  angles  (where   most  serendipitous   sources  are
detected), together  with the coarse  spatial resolution of  the \MOS\
cameras (1\farcs1/pixel),  would make the determination  of the source
extension  uncertain.    In  order   to  estimate  a   sky  coordinate
uncertainty  for  all  the   detected  sources,  we  recomputed  their
positions  using  the  task  {\tt emldetect},  which  performs  maximum
likelihood  fits to the  source spatial  count distribution.   In this
case,  we fixed  the threshold  values of  the equivalent  single band
detection  likelihood (parameter  \textit{mlmin}) to  30, in  order to
select only high--confidence sources.

Our  master lists contain  a total  of 132  sources for  the 2001/2002
observation subset, 107  sources for the 2005 subset,  and 144 sources
for  the whole  observation  set.  Although  we  performed the  source
detection using the  same tasks, the master list  presented in Paper I
contained 196 sources for  the 2002 observations only.  The difference
between the  number of sources in the  two lists is mainly  due to the
improvement of  the task  {\tt eboxdetect} in  the {\em  SAS} v.7.1.0,
which was used  to perform the source detection.   The task now allows
for  a more  accurate analysis  in regions  of diffuse  emission, thus
reducing  the detection of  spurious sources.   Moreover, we
now applied tighter  selection criteria in order to  qualify an X--ray
source as real.

\begin{figure*}
\begin{center}
\begin{tabular}{c@{\hspace{1pc}}c}
\includegraphics[height=8.5cm,angle=0]{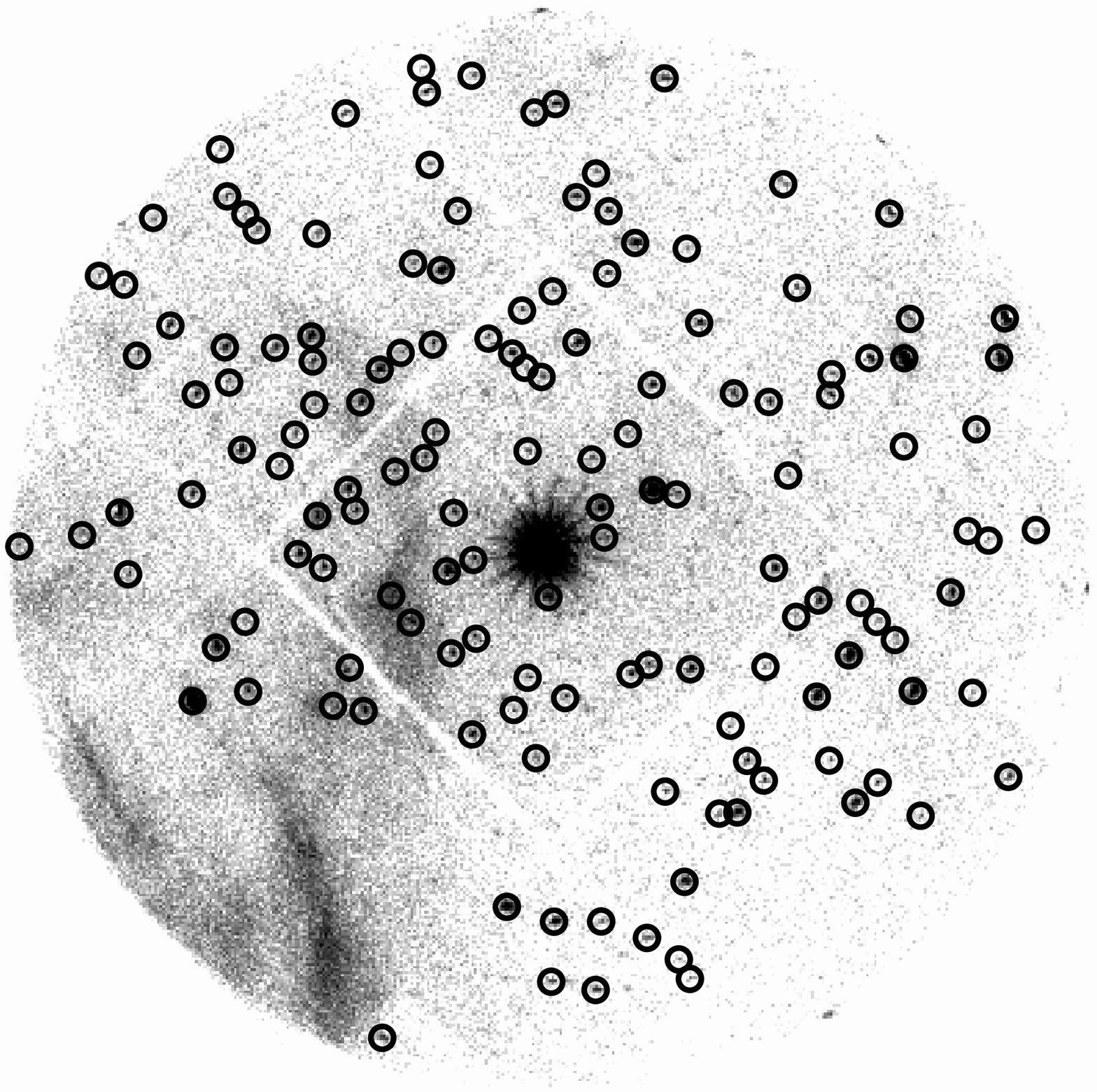} &
\includegraphics[height=8.5cm,angle=0]{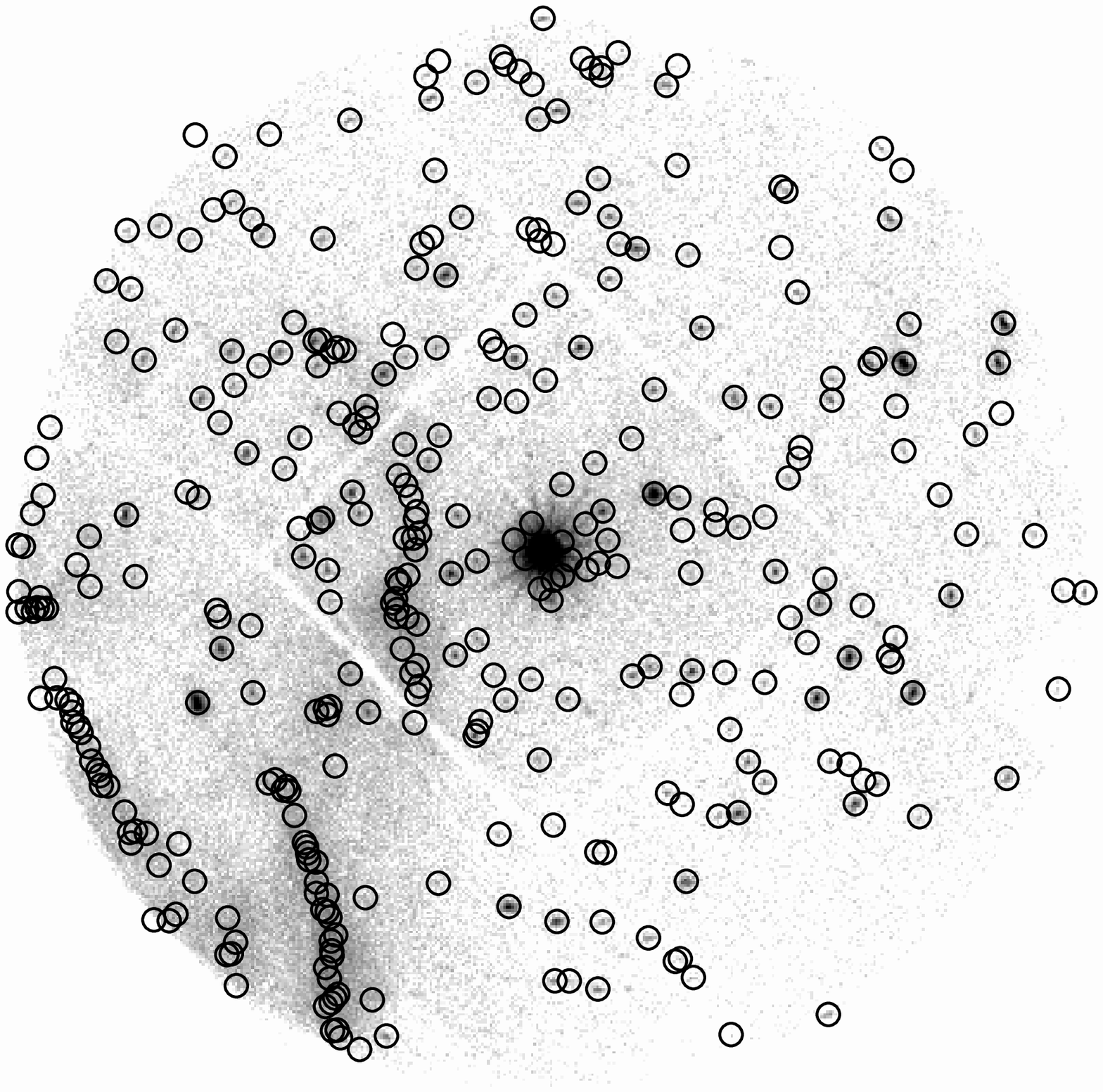} \\
\end{tabular}
\end{center}
\caption{\textit{Left:} distribution of the 144 X--ray sources detected in the \EPIC/{\MOS} image of \1E\ in the energy range 0.3--8 keV. \textit{Right:} distribution of the 344 X--ray sources listed in the \textit{2XMMi} catalogue in the same sky region.}\label{comparison2XMM}.
\vspace{-0.5 truecm}
\end{figure*}

We  note that  in the  field of  \1E\  the \textit{Incremental
Second            XMM--Newton           Serendipitous           Source
Catalogue\footnote{http://xmmssc-www.star.le.ac.uk/Catalogue/xcat\_public\_2XMMi.html}}
(\textit{2XMMi}, \citet{Watson2008,Watson2009}) reports  344 sources against the
144  found  by  our  source  detection procedure.  We  attribute  this
discrepancy mainly  to the  difference in the  threshold value  of the
\textit{detection  likelihood}  used  in  our  procedure  and  in  the
procedure used  to produce the \textit{2XMMi} catalogue.  In our case,
the  detection  likelihood is  set  to  8.5  for the  \textit{likemin}
parameter  of  the \SAS\  task  {\tt eboxdetect}  and  to  30 for  the
\textit{mlmin} parameter  of the \SAS~task {\tt  emldetect}, while for
the generation  of the \textit{2XMMi} catalogue  these parameters were
set               to               5              and               6,
respectively\footnote{http://xmmssc-www.star.le.ac.uk/Catalogue/UserGuide\_xmmcat.html}. In
other words, we applied much tighter criteria to the source selection,
thus rejecting several low--confidence  or spurious sources which are,
instead, included  in the \textit{2XMMi} catalogue. This  is proven by
Fig.~\ref{comparison2XMM}, where the sources detected by our procedure
are compared  with the \textit{2XMMi}  ones. As is  apparent, while
all our  sources have a \textit{2XMMi} counterpart,  the vast majority
of  the additional \textit{2XMMi}  sources are  either very  faint, or
detected at the edge of  the \textit{field--of--view}, or in region of
diffuse X--ray emission.  Therefore, it is quite likely 
that a large fraction of these sources are actually spurious.

In order to perform a  detailed statistical analysis we also computed,
for all the  observation sets, the number of  sources detected in each
of  the three   energy   bands.    We   summarised   these   numbers   in
Table~\ref{tab:Nsrc}  where we also  reported their  relative fraction
with respect to  the total number of sources detected  in at least one
energy  band.  We note  that almost  all sources  are detected  in the
total  energy band (0.3--8  keV), with  a good  fraction of  them also
detected in the  soft energy band (0.5--2 keV).  The number of sources
detected in each energy band is different across the three observation
sets, which is an effect of the uneven effective exposure times.  This
is  evident  in   the  case  of  the  2005   observation  subset  (see
Table~\ref{tab:expo}).

\begin{table}[h]
\begin{center}
\caption[]{Number of X--ray sources detected in each energy band and relative fraction for the three observation sets defined in \S~\ref{SourceDet}.} 
\label{tab:Nsrc}
\footnotesize{
\begin{tabular}[c]{c|ccc|c} \hline \hline
Band (keV) & 0.5--2 & 2--10 & 0.3--8 & Total \cr \hline
Set & N(\%) & N(\%) & N(\%) & N \cr \hline
 1 & 101 (76.5) & 68 (51.5) & 123 (93) & 132   \cr
 2 & 84 (78.5) & 42 (39) & 97 (90.6) & 107   \cr
 3 & 114 (72) & 87 (60) & 135 (94) & 144   \cr
\hline \hline
\end{tabular}}
\end{center}
\end{table}

\section{The serendipitous X--ray source catalogue}\label{sec:3}

\subsection{Catalogue description}

We used the source master list obtained from the whole observation set
to build a detailed catalogue of serendipitous X--ray sources detected
in  the \1E\ field.   The complete  serendipitous source  catalogue is
made available  in electronic form  through the {\em  Vizier} database
server.  Each  source  was  assigned  a unique  identifier  using  the
recommended  \XMM\   designations  for  serendipitous   sources.   The
catalogue information include most  of the parameters already included in
the  master list,  i.e.  sky  coordinates and  associated uncertainty,
effective exposure  time, total  counts, count--rate and  errors, {\em
S/N}  ratio,  and  detection  likelihood.  In  addition,  we  provided
information on the source  spectral parameters and the computed fluxes
in the soft/hard/total energy bands.

Since  for most sources  the measured  counts are  too few  to produce
significant  X--ray spectra, we  used the  {\em Hardness  Ratio} ({\em
HR}) to  provide qualitative spectral  information.  The {\em  HR} was
computed from the measured count--rate ({\em CR}) in the hard
(2--10  keV)  and  soft  (0.5--2  keV) energy  bands  and  is  defined
according to the equation:

\begin{equation}
HR=\frac{CR(2-10)-CR(0.5-2)}{CR(2-10)+CR(0.5-2)}
\end{equation}

where  $CR(2-10)$   and  $CR(0.5-2)$  are  the   count--rates  in  the
hard and soft energy bands, respectively.  The source flux in
the  soft/hard/total  energy  bands  was computed  from  the  measured
count--rates.  Following the  procedure used by \citet{Baldi2002}, the
{\em  count--rate--to--flux}   conversion  factors  ({\em   CF})  were
computed  for  each of  the  \MOS\  cameras  individually using  their
updated response matrices, combined  with the effective exposure times
of each pointing. As a  model spectrum we assumed an absorbed
power--law  with  photon  index  $\Gamma=1.7$,  i.e.   a  typical  AGN
spectrum,   and a hydrogen   column    density   $N_{H}$
=1.3$\times$10$^{21}$ cm$^{-2}$,  i.e. the value measured in
the direction of \1E.

In the following sub--section, we  report basic statistics on the more
important catalogue parameters, like the source {\em S/N} ratio,
the  total  {\em  CR},  and  the  {\em  HR}  relative  to  the  whole
observation  set.   In  the  last  sub--section we  also  present  the
log$N$--log$S$ distribution built  from the  sources  in our
serendipitous catalogue.

\subsection{Catalogue statistics}

\begin{figure}[!h]
\includegraphics[width=4.6cm,angle=-90,clip=]{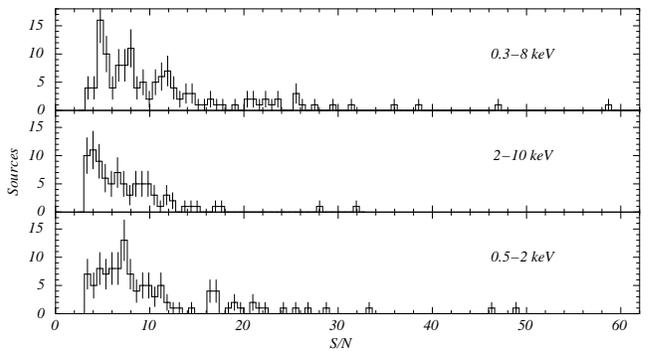}
\caption{Histogram of the {\em  S/N} ratio distribution in the energy bands 0.3--8 keV, 2--10 keV, and 
0.5--2 keV (top to bottom) for the serendipitous X--ray sources.}\label{histo_n_snr}
\vspace{-0.5 truecm}
\end{figure}

The histogram of the  source {\em signal--to--noise} ({\em S/N}) ratio
distribution is shown in Fig.~\ref{histo_n_snr} in the soft, hard, and
total  energy  bands.  In  the  total  energy  band (0.3--8  keV)  the
distribution  peaks  at  {\em  S/N}  =  4--6  (Fig.~\ref{histo_n_snr},
top). However, thanks  to the long effective integration  time and to
the increased  count statistics,  a large fraction  ($\sim$ 40  \%) of
sources are also  detected with {\em S/N} $\geq$  10.  Very few sources
are detected with {\em S/N} $\geq$ 20.  In the hard energy band (2--10
keV)  sources are  generally  detected  with a  quite  low {\em  S/N}
ratio,   with   the   peak   of   the   distribution   at   4
(Fig.~\ref{histo_n_snr},  middle) and with  only $\sim$  20 \%  of the
sources detected with {\em S/N} $\geq$ 10.  On the other hand, sources
are detected with the best {\em S/N} ratio in the soft energy
band   (0.5--2   keV),  with   the   distribution   peaking  at   6--8
(Fig.~\ref{histo_n_snr}, bottom),  and with a much  larger fraction of
sources ($\sim$ 35 \%) detected with {\em S/N} $\geq$ 10. This is most
likely ascribed to the better  sensitivity of the \MOS\ cameras at low
energies.

\begin{figure}[!h]
\includegraphics[width=6cm,angle=-90]{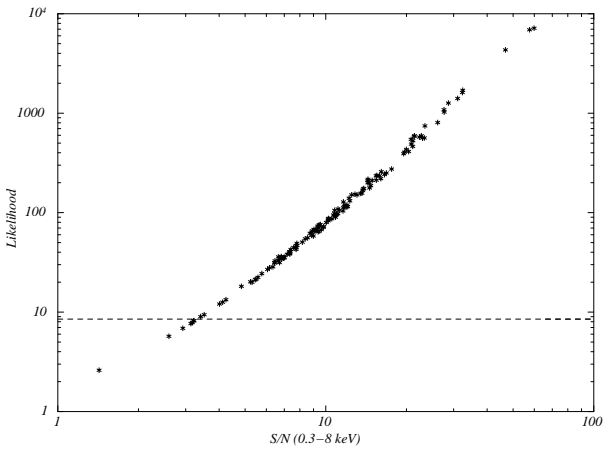}
\caption{Correlation  between the  detection likelihood  and  the {\em
S/N} ratio,  both  computed   in  the  0.3--8  keV  energy   band,  for  the
serendipitous sources.  The dashed  line corresponds to  the detection
likelihood threshold ($-ln  P = 8.5$) in the  0.3--8 keV band. Sources
below  this line  are included  because they  are above  the detection
threshold  in  at  least  one  of  the other  two  energy  bands  (see
\S~\ref{SourceList}). }\label{like_snr}
\vspace{-0.5 truecm}
\end{figure}

In  Fig.~\ref{like_snr}  we  show,  as a  reference,  the  correlation
between the  source detection likelihood  $-ln P$ and the  source {\em
S/N} ratio in the total energy band 0.3--8 keV.  As expected,
the detection  likelihood increases with the {\em  S/N} ratio, without
any  large scatter  or change  of  slope at  the two  extremes of  the
distribution.  

\begin{figure}[!h]
\includegraphics[width=3.4cm,angle=-90]{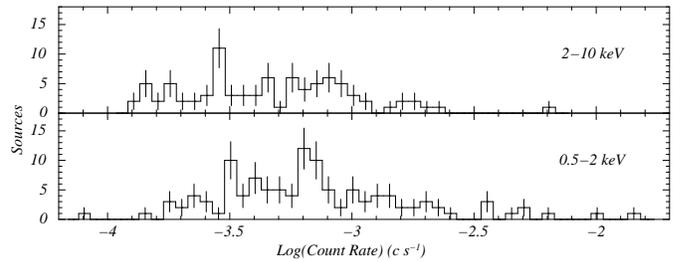}
\caption{Histogram  of  the  source  count--rate distribution  in  the
2--10   keV   and  0.5--2   keV   energy   bands   (top  and   bottom,
respectively).}\label{cr_n}
\vspace{-0.5 truecm}
\end{figure}

The histograms of the source {\em count--rate} ({\em CR}) distribution
for the two coarse soft (0.5--2 keV) and hard (2--10 keV) energy bands
are  shown  in  Fig.~\ref{cr_n}. The  peak of  the  {\em  CR}
distribution is  at 6.31 cts s$^{-1}$  ($logCR$ = -3.2)  and 2.81 cts
s$^{-1}$  ($logCR$  =  -3.55)  in  the soft  and  hard  energy  bands,
respectively. As seen  from the histograms, only a  few X--ray sources
have a  relatively high count--rate  ($logCR\geq -3$), and  thus lower
statistical  errors, in  either  of  the two  energy  bands. For  this
reason,  only these sources with count--rate  variations measured  over the
2001/2002 and 2005 observation subsets can be considered indicative of
a     statistically      significant     long     term     variability
(\S~\ref{sec:variab33}).

\begin{figure}[!h]
\includegraphics[width=6cm,angle=-90]{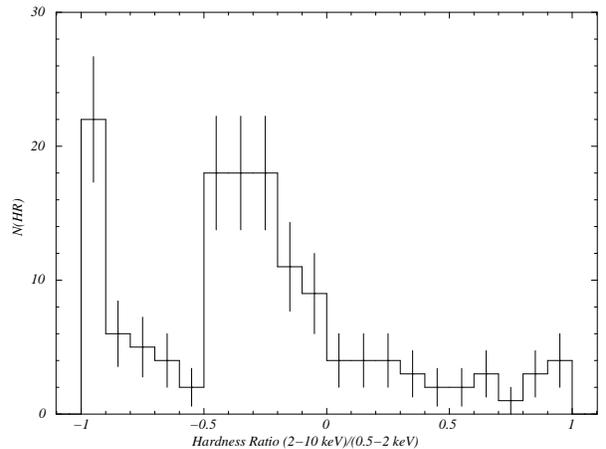}
\caption{Histogram of the {\em  HR} distribution for the serendipitous
X--ray sources  detected in the \1E\  field. 
}\label{n_hr}
\vspace{-0.5 truecm}
\end{figure}

The   histogram   of  the   {\em   HR}   distribution   is  shown   in
Fig.~\ref{n_hr}.  Most of  the sources have -0.5$\leq HR  \leq$0 and a
large  fraction has  $HR \sim  -1$. This  suggests that  a significant
fraction of the X--ray  source population is characterised by
rather soft spectra,  with no detection in the  2--10 keV energy band.
On the other hand, the histogram shows that only few sources have very
hard spectra ($HR \simeq 1$).

\subsection{Flux limit and sky coverage}

\begin{figure}[!h]
\includegraphics[width=6.5cm,angle=-90]{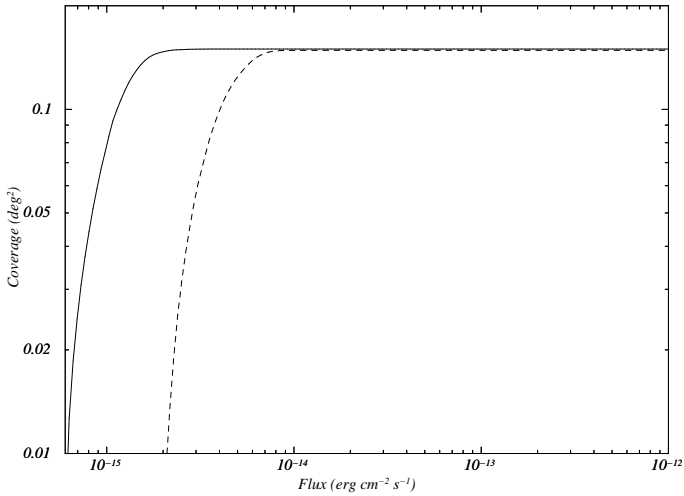}
\caption{Sky coverage  of the \XMM\  observations, in the  soft energy
range 0.5--2 keV ({\em solid line}) and in the hard energy range 2--10
keV ({\em dashed line}).}\label{cover}
\vspace{-0.5 truecm}
\end{figure}

The  actual sky  coverage  in the  various  energy ranges  was
computed by applying the  procedure described in \citet{Baldi2002}, which
is   consistent  with   the  standard   method  used   in   the 
\XMM~Serendipitous Survey \citep{Carrera2007,Mateos2008}. For
each energy  band we used the exposure  maps of each of  the two \MOS~
cameras and the  background map of the co--added  image, as derived in
\S~\ref{SourceDet}, to compute the flux limit map of the whole observation
set.   
To  this  aim,  we  applied  the  {\em  count--rate--to--flux}
conversion factors  ({\em CF})  obtained with the  absorbed power--law
spectrum described in  \S3.1. This gives, for each  position on the sky
covered by the \MOS\ observations, the flux that a source must have in
order   to   be   detected   with   a  minimum   probability   $P$   =
2$\times$10$^{-4}$  \citep{Baldi2002,Novara2006}.  We  used  the  flux
limit   maps   to   derive   the   total   sky   coverage   shown   in
Fig.~\ref{cover}. This shows that our observations cover a sky
area  of $\simeq$  0.15 deg$^{2}$,  down to  X--ray fluxes  of $\simeq
2\times10^{-15}$ and $8\times10^{-15}$  erg cm$^{-2}$ s$^{-1}$ for the
energy ranges 0.5-2 and 2-10 keV, respectively.

\subsection{LogN--logS distribution}

\begin{figure}[!h]
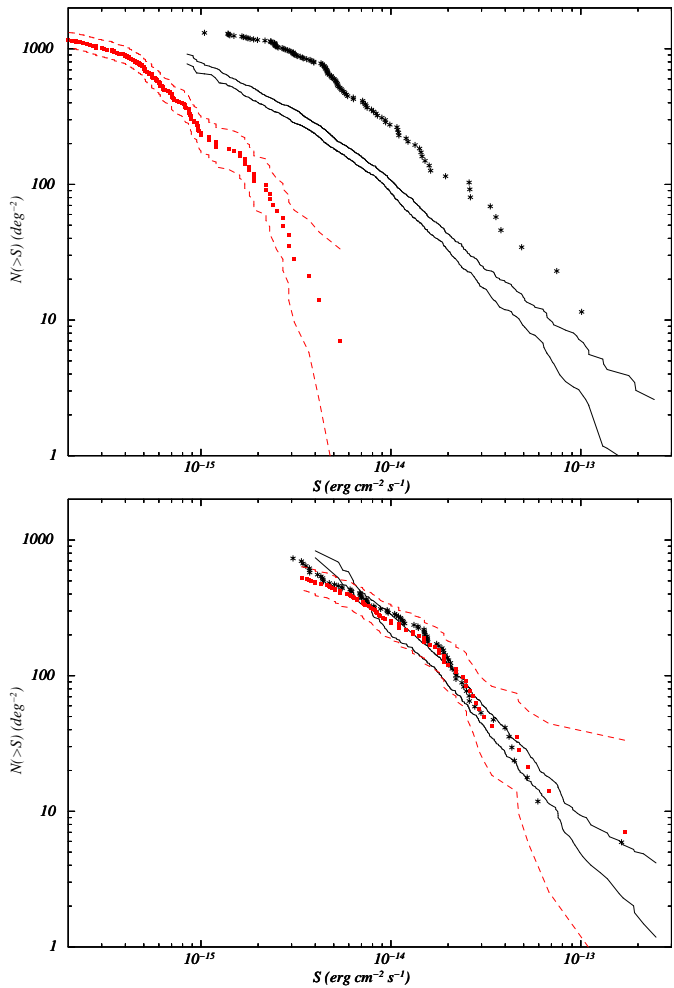

\includegraphics[width=6.5cm,angle=-90]{lognlogs_b1_all.ps} \\
\includegraphics[width=6.5cm,angle=-90]{lognlogs_b2_all.ps}
\caption{Cumulative log$N$--log$S$  distributions of the serendipitous
sources detected in  the \1E\ field in the soft  (0.5--2 keV, \textit{top}) and hard
(2--10 keV, \textit{bottom}) energy ranges  (\textit{asterisks}). The black solid lines mark the upper and lower limits obtained by
\citet{Baldi2002} in the same energy ranges but at higher galactic latitudes. The red filled squares and the red
dashed lines represent the distributions and the limits measured by {\em Chandra} in
the Galactic plane \citep{Ebisawa2005}, respectively.}\label{lnls_both}
\vspace{-0.5 truecm}
\end{figure}

We followed the procedure used by \citet{Baldi2002} to compute
the  log$N$--log$S$ distribution built  from our  serendipitous X--ray
sources,  and   we  refer  to   their  paper  for   further  details.
Fig.~\ref{lnls_both} shows  the cumulative log$N$--log$S$ distribution
(asterisks) relative  to the  soft (0.5--2 keV)  and hard  (2--10 keV)
energy bands (top and bottom panel, respectively). In the soft
band  the  flux  limit  is  $\sim 1  \times  10^{-15}$  erg  cm$^{-2}$
s$^{-1}$,  corresponding to a  maximum source  density of  $\sim 1300$
sources deg$^{-2}$, while in the hard energy band it is $\sim 3 \times
10^{-15}$ erg cm$^{-2}$ s$^{-1}$, corresponding to a source density of
$\sim 700$  sources deg$^{-2}$. Both  the soft and  hard distributions
feature an   evident   change   of   slope   at   $S   \sim
4\times$10$^{-15}$    and     $S    \sim    2\times$10$^{-14}$    erg
cm$^{-2}$s$^{-1}$,  respectively.   We  note that  a  similar
turn--over  was already  observed by  \citet{Ebisawa2005} in  the {\em
Chandra} observation  of the galactic  plane, and also  in the
the \XMM~\textit{Serendipitous Survey}
\citep{Carrera2007,Mateos2008}, even if in  the latter case the flux
breaks are at $S \sim 1\times$10$^{-14}$ erg cm$^{-2}$s$^{-1}$ in both
energy   bands.    With   respect   to   the   results   reported   by
\citet{Mateos2008},  we obtain a  comparable flux  limit in  the soft
energy  band,  while  in  the   hard  band  we  obtain  a  much  lower
limit. Moreover,  in both energy ranges our  cumulative source density
is  higher,  since they  obtain  $\sim  600$  and $\sim  300$  sources
deg$^{-2}$ in the  soft and hard energy ranges,  respectively. We note
that  we used a  unique power--law  spectral index  $\Gamma$ =  1.7 to
calculate   our  \textit{CFs}   in  the   two  energy   ranges,  while
\citet{Mateos2008}  used spectral indexes  of 1.9  and 1.6  below and
above 2 keV, respectively. However, they showed that differences $\le$
0.3 in the spectral  index can imply variations in the log$N$--log$S$ of only 1--2 \%  and of  $\le$  9 \% in the
hard and soft  bands, respectively.  Therefore, we assume
that  our results  are not  biased  by the  used spectral  parameters

For comparison, in Fig.~\ref{lnls_both} we superimposed on our
data  the lower  and upper  limits of  the log$N$--log$S$  computed by
\citet{Baldi2002}  at high  galactic  latitude ($|b|>27^{\circ}$).  We
note that,  with respect  to our work,  they obtained the  upper limit
log$N$--log$S$  by  applying  the  same detection  threshold  ($P_{\rm
th}=2\times10^{-4}$) but  a larger extraction radius,  while the lower
limit log$N$--log$S$ was obtained  with the same extraction radius but
a more constraining  threshold value ($P_{\rm th}=2\times10^{-5}$).  In
addition,  we overplotted  the  log$N$--log$S$ distributions  computed
from    {\em   Chandra}   observations    of   the    galactic   plane
\citep{Ebisawa2005}, as well as their 90 \% confidence limits.

In  the soft  energy band  our log$N$--log$S$  distribution is
well above  the high--latitude upper limit  of \citet{Baldi2002}. This
means that in  our serendipitous survey we detected  a large sample of
galactic sources  which are  missed not only  at higher  latitudes but
also in  the Galactic  plane, due to  the high amount  of interstellar
absorption.  However,  we note  that  our log$N$--log$S$  distribution
flattens  at low X--ray  fluxes with  respect to,  e.g. that  shown in
Paper  I,  with a  clear  break  at $S  \sim  4  \times 10^{-15}$  erg
cm$^{-2}$ s$^{-1}$.  This trend is due to the tighter criteria (see \S
2.4)  that  we adopted  to  validate  the  detection of  serendipitous
sources, together  with the  improved \SAS\ detection  algorithm which
minimises  the  number of  spurious  sources  detected  in regions  of
diffuse   emission,   like  those   associated   with   the  SNR   PKS
1209$-$51. This results  in a lower number of  sources detected in the
soft energy band, which is now 114 with respect to the 135 reported in
Paper I.  Indeed, we identified  the missing sources with the faintest
ones reported in Paper I, which explains the reduced number of sources
at  the low  flux  end  of the  new  log$N$--log$S$ distribution.  Our
log$N$--log$S$  distribution is  also  well above  the Galactic  plane
log$N$--log$S$ distribution (the  red points in Fig.~\ref{lnls_both}),
which means that we detected a significant fraction of extra--galactic
sources which are missed at low galactic latitude.

In the  hard energy  band, our log$N$--log$S$  distribution is
very close to that observed in the Galactic plane. With respect to the
high latitude  limits, our distribution  shows a slight excess  in the
flux range 1--2$\times 10^{-14}$  erg cm$^{-2}$ s$^{-1}$, possibly due
to the contribution of a fraction of Galactic sources which are missed
at  higher latitudes.  On  the other  hand,  the faintest  end of  our
distribution is below the high latitude lower limit. We attribute this
result to  the tight  criteria that we  used to validate  the detected
source, which implies the rejection of the faintest objects.

\section{The bright source sample}\label{sec:4}

\subsection{Spectral analysis}

\begin{figure}[!h]
\includegraphics[width=9cm,angle=0]{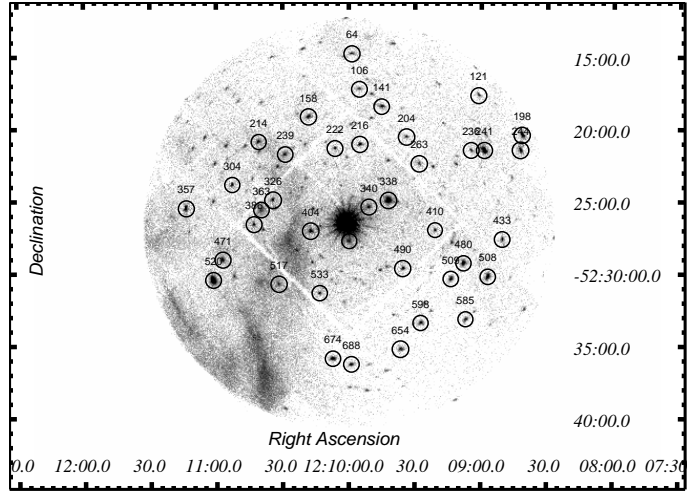}
\caption{Processed \XMM\ \textit{EPIC/MOS}  0.3--8 keV band image  of the \1E\
field  with the  position of  the 40  brightest  serendipitous sources
over plotted. Sources  are labelled according  to the numbering  used in
Table~\ref{sources}.  Circles are drawn  only to highlight  the X--ray
source  positions and  their size  do not  correspond to  their actual
positional errors. }\label{ima_x}
\vspace{-0.5 truecm}
\end{figure}

\begin{figure}[!h]
\includegraphics[width=5cm,angle=-90]{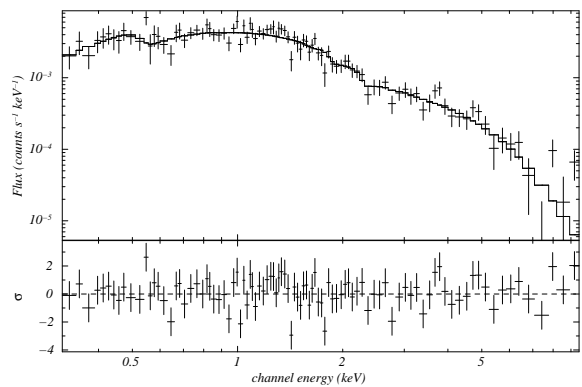}
\caption{Unbinned   non--thermal  spectrum   of  source   \#338  (XMMU
J120942.1$-$522458)  with the  best--fit {\em  power--law} model. 
}\label{spec338}
\vspace{-0.5 truecm}
\end{figure}

\begin{figure}[!h]
\includegraphics[width=5cm,angle=-90]{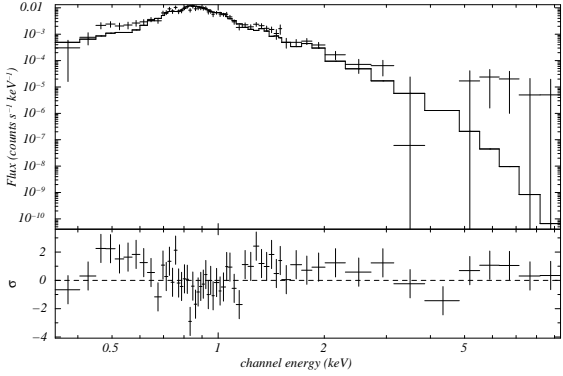}
\caption{Unbinned   thermal    spectrum   of   source    \#241   (XMMU
J120858.8$-$522129) with the best--fit {\em thermal mekal} model. 
}\label{spec241}
\vspace{-0.5 truecm}
\end{figure}

Although the {\em HR} provides qualitative information on the source
X--ray spectra, it is not a robust spectral
classification.  As  we mentioned  in Paper  I, at
least   500   total  \MOS\   counts   (i.e.  \textit{MOS1}   +
\textit{MOS2}  events)  over  the  whole detector  energy  range  are
required  to  discriminate thermal  X--ray  spectra from  non--thermal
ones.  Following this criterion,  we selected the 40 brightest sources
(Fig.~\ref{ima_x}) in  our serendipitous source  catalogue which total
$>$ 500 counts.   This bright source sample obviously  includes the 24
brightest sources similarly selected  in Paper I. For each of
the  two \MOS\  cameras we  extracted the  source  event list  using
extraction  radii of  20\arcsec--35\arcsec.   Background regions  were
selected near  the source  positions, with a  radius three  times that
used for the source extraction.   All spectra extracted from the event
lists were rebinned in order to have a minimum of 30 counts per energy
bin, which is required  to precisely apply the $\chi^{2}$ minimization
fitting  technique. For  each  of the  two  \MOS\ spectra we
generated {\em ad hoc} response matrices and ancillary files using the
\SAS\ tasks  {\tt rmfgen} and {\tt  arfgen}  
with both  thermal and non--thermal spectral  models. We took
into  account  the  different   size  of  the  source  and  background
extraction areas and renormalized  the background count--rate, then we
simultaneously fitted the two spectra  of each source, forcing common
parameters  and allowing  only for  a cross--normalization  factor to
account  for  the different  instrument  efficiency.  We
considered four    spectral   models:   {\em    power--law},   {\em
bremsstrahlung}, {\em  black--body}, and  {\em mekal}.  In  all cases,
the hydrogen column density $N_{H}$  was left as a free parameter. For
each emission  model we computed the  90 \% confidence  level error on
both  the $N_{H}$  and on  the spectral  parameters, i.e.   the plasma
temperature or  the photon--index.  As  seen from Table~\ref{sources},
we found that 14 sources were  best fitted by a {\em power--law} model
(Fig.~\ref{spec338}), 2  by a {\em  bremsstrahlung} model, and 3  by a
{\em mekal}  model (Fig.~\ref{spec241}).  For  16 of the  remaining 20
sources, at least two different models provided an acceptable fit with
a comparable  value of the $\chi_{\nu}^{2}$. For  5 sources it
was not possible to  obtain acceptable results with single--component
spectral  model.   This is,  e.g.   the  case  for source  \#239  (XMMU
J121029.0$-$522148),  the  proposed  Seyfert--2 galaxy  identified  in
paper I, which is characterised  by a complex spectral model as discussed
in \S~\ref{sec:sy2}.

\subsection{Time variability}\label{sec:variab33}

\begin{figure}[!h]
\includegraphics[width=6cm,angle=-90]{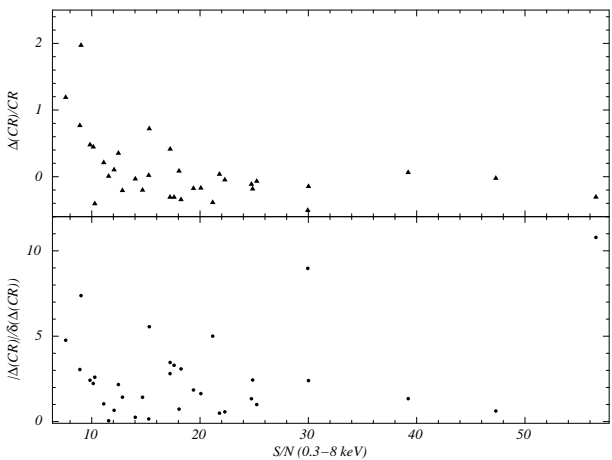}
\caption{Top: relative  {\em CR}  variation with  respect to  the  first observation
subset plotted as a function of the source {\em S/N} ratio (filled triangles). Bottom: absolute {\em  CR} variation $|\Delta  CR|$ divided by its associated error  $\delta (\Delta  CR)$ plotted as a function of the
source {\em S/N} ratio (filled circles).}\label{macrovar}
\vspace{-0.5 truecm}
\end{figure}

In  order to investigate  possible long  term variability between exposures we selected
from  our bright source sample 33  X--ray  sources that  we
detected  in both  the  2001/2002 and  2005  observation subsets  (see
Table~\ref{tab:expo})  and  in the  total  (0.3--8  keV) energy  band,
chosen as  a reference.  For each source  we computed  the count--rate
variation   $\Delta   CR$  between   the   two  observation   subsets.
Fig.~\ref{macrovar} (top)  shows the relative {\em  CR} variation with
respect to the  first observation subset plotted as  a function of the
source {\em S/N} ratio.  As seen,  a few sources show non--zero long
term variability which  is mostly within 30\% but can  be up to $\sim$
200  \%.  Fig.~\ref{macrovar} (bottom)  shows  the  absolute {\em  CR}
variation  $|\Delta  CR|$  divided  by its  associated  error  $\delta
(\Delta CR)$ plotted as a function  of the source {\em S/N} ratio.  As
seen, 10  X--ray sources show evidence  of variability at  more than 3
$\sigma$.    For  6   of  them, i.e. source   \#326  (XMMU
J121034.6$-$522457),  \#404  (XMMU  J121017.5$-$522706),  \#410  (XMMU
J120921.0$-$522700),  \#471  (XMMU  J121057.3$-$522905),  \#480  (XMMU
J120908.1$-$522918),   and   \#520   (XMMU  J121101.5$-$523030),   the
variability is at the \gtsima\ 5 $\sigma$ level. We thus regards these
sources  as  likely  transients. Among them,  source \#326  (XMMU J121034.6$-$522457)
features  the  strongest  variability  ($\sim$  200  \%,  $\approx$  7
$\sigma$),  followed by source  \#410 (XMMU  J120921.0$-$522700) whose
variability is of $\sim$ 100 \%  but is significant only at the $\sim$
5   $\sigma$  level.   On  the   other  hand,   sources   \#480  (XMMU
J120908.1$-$522918)  and  \#520  (XMMU J121101.5$-$523030)  feature  a
variability  of only  $\sim$  15--25 \%,  although  detected with  the
highest significance  ($\approx 10 \sigma$). This is  obviously due to
the fact  that both sources were  detected with the  highest {\em S/N}
ratio ($\ge$ 30).

For all  sources in our bright  sample, we searched
for  variability  on  a  shorter time
scale, including within exposures, through a  light--curve
analysis with optimised  time binning (1, 5, or 10 ks)  and using as a
reference only {\em CR} measurements relative to the observations with
the longest exposure times, i.e.  the second and the third observation
of the 2002 data--set and the  fourth observation of the 2005 data set
(Table~\ref{tab:expo}). Of the 6 X--ray sources with $\geq$ 5 $\sigma$
possible  long term  variability, our  light--curve analysis  does not
show evidence  of short  term variability while  it confirms  the long
term   one  for   all  sources   but  not   for  source   \#520  (XMMU
J121101.5$-$523030).  This result is  not surprising since this source
is  the one  with the  lowest relative  variation ($\sim$  25  \%, see
Fig.~\ref{macrovar} (top)), which is  thus more difficult to recognise
if spread  on a  shorter time  scale. None of  the remaining  4 X--ray
sources  with possible  long term  variability ($\sim$  3--5 $\sigma$)
shows any evidence of short term variability.

For the persistent sources (long  term variability $\le 3 \sigma$), we
confirm flux variability on time--scales  of a few hundred seconds for
source \#158 (XMMU J121018.4$-$521911) and  of $\sim$ 10 ks for source
\#338 (XMMU  J120942.1$-$522458), which correspond to  source \#72 and
\#183 of Paper I, respectively.  For the remaining sources, we did not
find evidence of significant variability on any time scale.

Finally,  we  also  looked  for possible  periodic  time  variability.
Unfortunately,  in this case  the low  count statistics  prevented the
detection of  any periodic signal at a  reasonable significance level.

\subsection{The Seyfert--2 galaxy ESO 217-G29}\label{sec:sy2}

Source \#239 (XMMU J121029.0$-$522148)  was originally identified as a
new  Seyfert galaxy  in  Paper I  (source  \#127), due  to its  X--ray
spectrum  and  to  its  positional  coincidence with  the  galaxy  ESO
217-G29, a  bright ({\em R}=14.93) barred spiral  with a spectroscopic
redshift of  0.032 \citep{Visvanathan1992}  also detected in  the {\em
Digitised Sky  Survey} images.  From  the merged image (see  \S~\ref{SourceDet}) we
have now obtained a total of 821 counts in the energy range 0.3--8 keV
for source \#239, which is 38\% higher with respect
to that of the data set  used in Paper I.  For this reason, we
repeated  the source  spectral analysis  in  order to  achieve a  more
accurate characterisation of the X--ray spectrum.  The spectrum of the
source  between 1 and  12 keV  is complex  and cannot  be fitted  by a
single-component  model. We  thus used  the AGN  unification  model of
\citet{Antonucci1993} and \citet{Mushotzky1993}

\vspace{+0.25                                                       cm}
$S=A_{G}[A_{SP}(R_{W})+A_{T}(PL+R_{C}+GL)]$\footnote{{\tt
wabs*(zwabs*powerlaw +  zwabs*(powerlaw +  pexrav + zgauss))}  in {\em
XSPEC}} \vspace{+0.25 cm}
 
where  $A_{G}$  is   the  galactic  absorption  (1.28$\times$10$^{21}$
cm$^{-2}$), $A_{SP}$ is the absorption related to the AGN host galaxy,
$R_{W}$ is  the warm and optically thin  reflection component, $A_{T}$
is the  absorption acting  on the nuclear  emission associated  to the
torus of dust  around the AGN nucleus, $PL$  is the primary power--law
modelling  the nuclear component,  $R_{C}$ is  the cold  and optically
thick reflection  component, and $GL$  is the Gaussian  component that
models the  Fe line at 6.4  keV.  For the  $A_{SP}$, $A_{T}$, $R_{C}$,
and  $GL$  components the  redshift  value is  fixed  at  $z$ =  0.032
\citep{Visvanathan1992}.

\begin{figure}[!h]
\includegraphics[width=6cm,angle=-90]{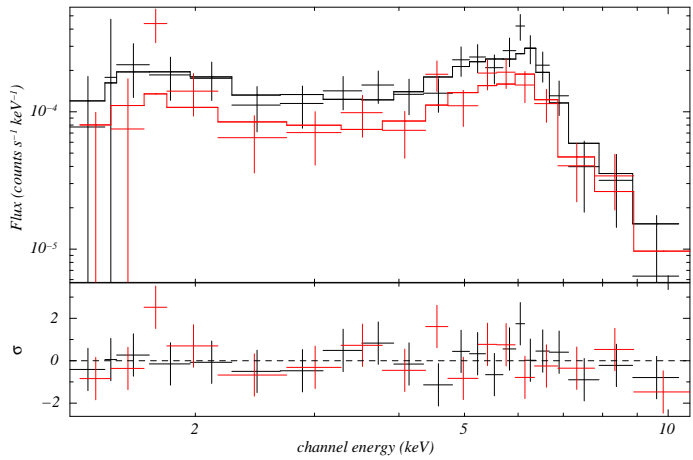}
\caption{(\textit{Upper panel}) The 1.2--12 keV unbinned spectrum of source \#239 (XMMU J121029.0$-$522148) identified with the Seyfert--2 galaxy ESO 217-G29. The fit was performed using  the      AGN      unification       model      of
\citet{Antonucci1993} and \citet{Mushotzky1993} with a fixed redshift of $z$ = 0.032. Spectral fits were computed for both the \textit{MOS1} and \textit{MOS2} data (black and red, respectively). (\textit{Lower panel}) Data--model residuals are shown in units of $\sigma$.}\label{zfix}
\vspace{-0.5 truecm}
\end{figure}

\begin{figure}[!h]
\includegraphics[width=6cm,angle=-90]{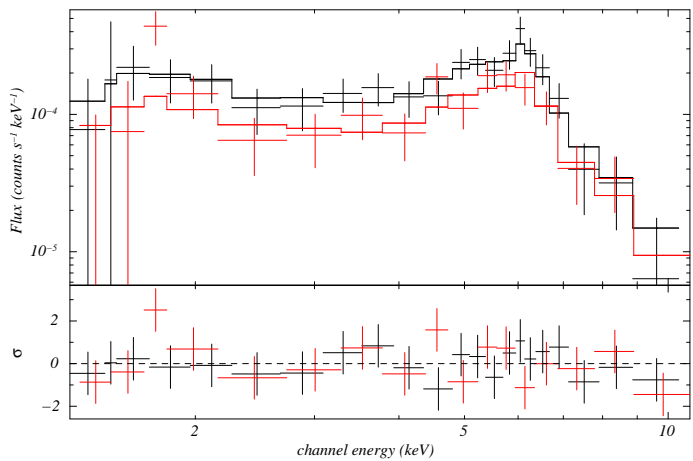}
\caption{Same as Fig.~\ref{zfix} but with the best-fit redshift value of $z$ = 0.042.}\label{zfree}
\vspace{-0.5 truecm}
\end{figure}

For both the \textit{MOS1}  and \textit{MOS2} spectra we performed the
spectral fitting both fixing the  redshift $z$ to the literature value
of  0.032 and  leaving it  as  a free  parameter.  In  the first  case
(Fig.~\ref{zfix}), the fit  yields a $\chi^{2}_{\nu}=0.77$ (33 d.o.f.)
but  it does  not satisfactorily  account for  the Fe  line  since the
fitted centroid  energy of the  line is 6.2  keV instead of 6  keV, as
actually measured  in the unfitted spectrum.   Furthermore, the fitted
line is not significant with  respect to the model continuum.  The fit
yields an absorption associated with the dust torus ($A_{T}$) of $\sim$
71.91$\times$10$^{22}$  cm$^{-2}$,  slightly   lower  than  the  value
reported in Paper  I.  In the second case  (Fig.~\ref{zfree}), the fit
also  yields  a  $\chi^{2}_{\nu}=0.77$  (32  d.o.f)  with  a  best-fit
redshift value  $z$ =  $0.042^{+0.038}_{-0.032}$ which is  between the
value reported  in Paper  I ($z$  = 0.057) and  the literature  one of
0.032.  The  fit with  the free  $z$ better accounts  for the  Fe line
whose fitted profile is now  significant at the 90\% confidence level,
with  a  fitted centroid  energy  of  $\sim$  6.0 keV.  The  intrinsic
absorption  associated   to  the   dust  torus  ($A_{T}$)   is  $\sim$
72.16$\times$10$^{22}$  cm$^{-2}$,   very  similar  to   the  previous
case. All best-fit parameters for  the two cases are summarised in the
Table~\ref{fit}.  The 2--10 keV  unabsorbed flux (calculated with {\em
XSPEC})      of     the      primary     nuclear      component     is
6.59$^{+2.13}_{-1.23}\times$10$^{-13}$ erg  cm$^{-2}$ s$^{-1}$ and the
X--ray   luminosity,   computed   for   a  redshift   of   0.032,   is
2.75$^{+0.89}_{-0.51}\times$10$^{42}$ erg s$^{-1}$.

\begin{table}[htbp]
\begin{center}
\caption{Best--fit parameters for source \#239 (XMMU J121029.0$-$522148), for the optical redshift $z$ = 0.032 and for its best--fit value $z$ = 0.042.}\label{fit}
\begin{tabular}{cccc} \hline \hline
Component 	& Parameter		& $z$=0.032 (fix)			& $z$=0.042			\\ \hline
$A_{SP}$		& N$_{\rm H1}^{a}$	& 2.27$_{-0.88}^{+1.12}$		& 2.27$_{-0.82}^{+1.12}$		\\
$R_{W}$		& $\Gamma$		& 1.9 (fixed)			& 1.9 (fixed)			\\
		& Flux @ 1 keV$^{b}$	& 8.84$_{-2.65}^{+3.25}$		& 8.63$_{-1.87}^{+1.58}$		\\
$A_{T}$		& N$_{\rm H2}^{a}$	& 71.91$_{-15.24}^{+16.18}$	& 72.16$_{-15.20}^{+20.92}$	\\
$PL$		& $\Gamma$		& 1.9 (fixed)			& 1.9 (fixed)			\\
		& Flux @ 1 keV$^{c}$	& 2.28$_{-0.79}^{+1.17}$		& 2.20$_{-0.41}^{+0.71}$		\\
$R_{C}$		& $\Gamma$		& 1.9 (fixed)			& 1.9 (fixed)			\\
		& Flux @ 1 keV$^{c}$	& 2.28$_{-0.79}^{+1.17}$		& 2.20$_{-0.41}^{+0.71}$		\\
$GL$		& $E_{line}$(keV)		& 6.4 (fixed)			& 6.4 (fixed)			\\
		& $I_{line}^{d}$		& 1.02$_{-1.02}^{+1.32}$		& 1.31$_{-1.26}^{+1.45}$		\\
		& EQW (eV)		& 126$_{-126}^{+164}$		& 164$_{-158}^{+181}$		\\ \hline
d.o.f.		&			& 33				& 32				\\
$\chi^{2}_{\nu}$	&			& 0.77				& 0.77				\\ 
\hline \hline
\end{tabular}
\end{center}
\begin{small}

$^{a}$ $10^{22}$ cm$^{-2}$

$^{b}$ $10^{-6}$ ph cm$^{-2}$ s$^{-1}$ keV$^{-1}$

$^{c}$ $10^{-4}$ ph cm$^{-2}$ s$^{-1}$ keV$^{-1}$

$^{d}$ $10^{-6}$ ph cm$^{-2}$ s$^{-1}$
\end{small}
\end{table}

\section{Optical observations}\label{sec:5}

\subsection{Observation description}

\begin{figure}[!h]
\includegraphics[width=8cm,angle=0]{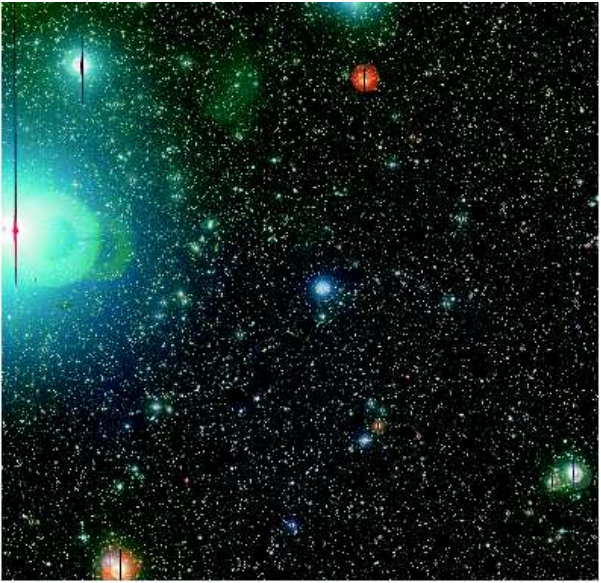}
\caption{Composite  {\em  VRI}  image  of the  \1E\  field  (34\arcmin
$\times$  34\arcmin) taken with  the {\em  WFI} at  the ESO/MPG  2.2 m
telescope. North  to the top,  east to the  left.  The effects  of the
very bright star $\rho$ Cen are clearly visible on the image, with the
presence of reflections and bright ghosts.}\label{ima_wifi}
\vspace{-0.5 truecm}
\end{figure}

In order to search for  the optical counterparts of the X-ray sources,
we  performed follow--up  observations (Fig.~\ref{ima_wifi})  with the
\WFI\  mounted  at  the  2.2  m  ESO/MPG telescope  at  the  La  Silla
observatory (Chile). The \WFI\ is a wide field mosaic camera, composed
of eight 2048$\times$4096 pixel CCDs, with a scale of 0\farcs238/pixel
and  a full field  of view  of 33\farcm7$\times$32\farcm7,  which well
matches that of the \EPIC/\MOS\ cameras.  Observations in the
{\em U}, {\em B}, {\em V}, {\em R}, and {\em I} filters were performed
in   Service   Mode  between   March   2005   and   April  2006   (see
Table~\ref{tab:wfi}).   Unfortunately,  scheduling problems  prevented
observations being executed during the same run. To compensate for the
inter chip  gaps, pointings were  split in sequences of  five dithered
exposures with shifts of  35\arcsec\ and 21\arcsec\ in right ascension
and declination,  respectively. The  target field was  always observed
close to the zenith and nearly always under sub-arcsecond seeing
conditions, as measured by the La Silla DIMM seeing monitor.

\begin{table}[h]
\begin{center}
\caption[]{Summary of the \WFI\ optical observations performed by the ESO/MPG 2.2m telescope.}\label{tab:wfi}
\footnotesize{
\begin{tabular}{l|crcc} \hline \hline
Date & Filter & Time (s) & Airmass  & Seeing \\ \hline 
10 Mar 2005 & R & 2888.75 & 1.09 & 0.99 \\ 
            & I & 1999.73 & 1.15 & 1.09 \\ 
01 May 2005 & V & 2559.09 & 1.11 & 0.52 \\  
02 May 2005 & I &  269.42 & 1.10 & 0.69 \\ 
25 Feb 2006 & I & 1999.59 & 1.09 & 0.86  \\ 
25 Apr 2006 & B & 1999.59 & 1.09 & 0.68 \\ 
            & U & 2499.59 & 1.11 & 0.61 \\
\hline \hline
\end{tabular}}
\end{center}
\end{table}

\subsection{Data reduction and calibration}

The data  reduction of the \WFI\  data was performed  with the \theli\
pipeline \citep{Erben2005}  which was also  used for the  reduction of
the \WFI\ data of  \citet{LaPalombara2006}. Since we followed the same
procedures,  we refer to  the paper  of \citet{LaPalombara2006}  for a
more detailed  description of the  data reduction.  Briefly,  for each
band  the  individual   images  were  de--biased,  flat--fielded,  and
corrected  for  the  fringing.   After  the  chip-by-chip  astrometric
calibration  (average  rms  $\sim
0\farcs3$) computed using a number of well-suited  (i.e., bright but not
saturated and not  detected close to the chip  edges) reference stars
selected  from  the  {\em  USNO-B1.0}  catalogue \citep{Monet2003}, single frames  were co--added  using a  weighted  mean to
reject cosmic ray hits.   A flux--renormalisation to the same relative
photometric zero-point was applied using the exposure maps produced by
the  pipeline  to account for the  uneven  exposure produced  by  the
dithering.  Since standard star observations were not acquired for all
nights    and    for    all    bands,   we    used    default    \WFI\
zero-points\footnote{http://www.ls.eso.org/lasilla/sciops/2p2/E2p2M/WFI}
for the photometric calibration, namely 21.96, 24.53, 24.12, 24.43 and
23.37 (in Vega magnitudes) for the {\em U}, {\em B}, {\em V}, {\em R},
and  {\em   I}  filters,  respectively.   A  deeper   image  was  then
constructed  by registering  the  individual co--added  images in  the
single bands, which was used as a reference for the source detection.

\subsection{Source detection}
 
The source extraction was performed on the final co--added single band
images  by running the  {\em SExtractor}  software \citep{Bertin1996}.
The source detection was performed after masking the region around the
very bright star $\rho$ Cen, a  B3V star (V=3.9) that was saturated on
all  \WFI\  images  (Fig.~\ref{ima_wifi}).  This  was  done  to  avoid
including spurious detections produced by the saturation spikes and to
filter out  objects whose  photometry is polluted  by the  bright star
halo.  The  masking was applied on  the weighted images and,  due to the
different brightness  of the  star in the  different bands and  to the
different integration time, the size of the masked region was tailored for each image.   
The extracted  catalogues were  checked  against the
images and the  counterparts were visually inspected to make sure that the
spurious detections were minimal (less than $\sim$ 1 \%).  Single band
optical  catalogues  were then  matched  using  a  matching radius  of
0\farcs2, i.e. equal to the rms of our astrometric solution, to
produce  the final \WFI\  colour catalogue.  The catalogue  includes a
total of  64910 sources with at least  a detection in one  of the five
bands ({\em  UBVRI}). Of these, only  15201 have been  detected in all
bands.   For  each  filter,  the  limiting  magnitude  of  the  colour
catalogue was defined as the magnitude of the object fainter than the
remaining  99\%.   This corresponds  to  {\em  U}-to-{\em I}  limiting
magnitudes of 23.25, 24.72, 24.39, 23.97 and 22.72.

\subsection{The optical/NIR catalogue}

To   extend  the   colour  coverage,   required  for  a  colour--based
classification  of the  \WFI\ sources,  we added  near  infrared (NIR)
photometry information  in the {\em J},  {\em H} and {\em  K} bands by
correlating  the \WFI\  colour  catalogue with  the \tmass\  catalogue
(Skrutskie  et al.  2006).  The  extracted  \tmass\ source  list in  a
$40\arcmin  \times  40\arcmin$ region  around  the  \1E\ position  was
retrieved through  the {\em Vizier}  database server and  matched with
the \WFI\ colour  catalogue using the IRAF task  {\tt tmatch}. A match
radius of 0\farcs5 was used to account both for the uncertainty on the
\WFI\ coordinates  and on the  $\le$ 0\farcs2 astrometric  accuracy of
\tmass. A total of 6996 \WFI\ sources ($\sim$ 10 \%) have a match with
a  \tmass\  source  and  for  5032  of them  we  have  the  full  {\em
UBVRI}-to-{\em  JHK}  photometry  information.  The match  produced  a
master  optical/NIR catalogue  that we  used  as a  reference for  the
X--ray   source  identification  and   for  the   colour-based  object
classification.  For  all sources with an  adequate colour-coverage we
used  the colour-based optical  classification technique  described in
\citet{Eva2002a}     and     tested     in    \citet{Eva2002b}     and
\citet{Groenewegen2002}.

\section{X--ray vs. optical/NIR catalogues}\label{sec:6}

\subsection{Catalogue cross--correlations}

In order to identify candidate  counterparts to the X--ray sources, we
cross-matched  our  serendipitous  X--ray  source catalogue  with  the
optical/NIR master catalogue.  Thanks  to the improved \SAS\ task {\tt
emldetect}, the  coordinates of the X--ray sources  were measured with
high accuracy. The measured errors vary between 0\farcs1 and 1\farcs5,
depending  on  the source  counts,  with  an  average error  of  $\sim
0\farcs7$. These errors, however, substantially reflect the positional
accuracy of the X--ray sources  with respect to the detector reference
frame and do not account  for systematic errors.  Indeed, the absolute
accuracy  of  these  coordinates  with respect  to  the  International
Celestial  Reference  Frame  (ICRF)  is  inevitably  affected  by  the
precision of the satellite aspect solution. In order to determine the accuracy
of  the  tie  of  the   measured  coordinates  to  the  ICRF,  we  thus
cross--matched  our  X--ray  catalogue  with  the  optical/NIR  master
catalogue.   In this way,  we found  six X--ray  sources which  have a
single,  relatively bright  (but  not saturated)  and obvious  optical
counterpart which is not  at the edges  of the \MOS\  cameras field--of--view
({\em FOV}).   We thus computed the linear  transformation between the
X--ray and optical coordinates to correct the \MOS\ astrometry.  Since
the  astrometry of the  \WFI\ catalogue  is calibrated  with USNOB-1.0, 
 which  is tied  to the ICRF,  we are  sure that we  did not
introduce a bias  in our procedure.  Using the  IRAF task {\tt geomap}
we found that the X--ray source coordinates are affected by a (radial)
systematic astrometric error of 1\farcs34, corresponding to the rms of
the X--ray--to--optical  coordinate transformation.\footnote{We note
that  in  Paper I  the  systematic  astrometric  error of  the  X--ray
coordinates was 2\farcs33, the  discrepancy being due to the different
counterpart assumed for one of  the six X--ray reference sources.}  To
this, we have  to add in quadrature the  measured positional statistic
error  of  each  source  (0\farcs1--1\farcs5).  Therefore,  the  total
uncertainty  on the X--ray  source position  is between  1\farcs34 and
2\farcs01.  The  correction to the X-ray coordinates  was then applied
to all  sources of  our serendipitous X--ray  catalogue with  the IRAF
task  {\tt   geoxytran}  using   the  coefficients  of   the  computed
X--ray--to--optical  coordinate  transformation.  To account  for  all
other sources  of uncertainty,  e.g.  the $\sim$ 0\farcs2 absolute 
accuracy (per coordinate) of the {\em USNO-B1.0} reference  frame  \citep{Monet2003},  
the
distortions of  the \MOS\ cameras,  etc., in the  X--ray--to--optical
cross--correlation we conservatively  assumed a more generous matching
radius equal to three times the estimated absolute error on the X--ray
source coordinates.

\subsection{Sources with candidate optical counterparts}

After applying the computed  correction to the coordinates of the
X--ray sources in our  serendipitous catalogue (see previous section),
we  repeated the cross--correlation  with the  optical/NIR catalogue.
After the cross--match we found at least one candidate counterpart for
112 out of the 144 X--ray sources in our serendipitous catalogue (i.e.
78 \% of  the total).  However, a total  of 195 candidate counterparts
were  found since  we  obtained multiple  matches  for several  X--ray
sources.  Due to the relatively  deep limiting magnitudes of the \WFI\
observations, this is in line  with the expectations.  We note that in
Paper I, where we used the shallower \gsc\ catalogue (with only $\sim$
16000  optical  sources instead  of  the  almost  65000 of  the  \WFI\
catalogue), we found at least one candidate counterpart only for about
half  of the  X--ray sources,  even  using a  more conservative  fixed
positional uncertainty, hence  a more generous cross--matching radius,
of 5\arcsec.  The choice of assuming a fixed positional uncertainty in
Paper I was  dictated by the fact that the  \SAS\ task {\tt emldetect}
was failing in providing reliable positional errors.

\begin{figure}
\includegraphics[width=6.5cm,angle=-90]{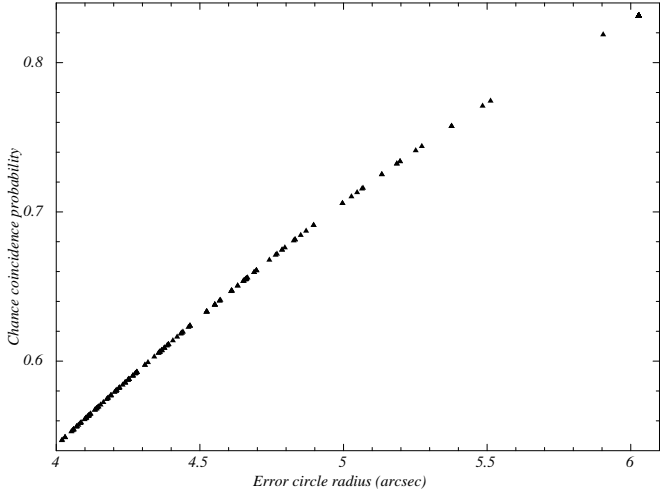}
\caption{Chance coincidence  probability between an  X--ray source and
an optical \WFI\  source as a function of  the matching radius assumed
equal  to three  times the  size  of the  estimated absolute  position
uncertainty.}\label{pcc}
\vspace{-0.5 truecm}
\end{figure}

Due to the contamination of fore/background objects, the result of the
cross--matching between the X--ray and optical catalogues is obviously
affected  by spurious  matches. In  order  to estimate  the number  of
spurious matches, we used the relation $P=1-e^{-\pi r^{2} \mu}$, where
$r$ is  the assumed  X--ray matching radius  and $\mu$ is  the surface
density per  square arcsecond of  the optical sources, to  compute the
chance coincidence probability between an X--ray and an optical source
\citep{Severgnini2004}.  In  our case, the \WFI\  catalogue provided a
total of 64910 sources distributed  over an area of about 34$\times$34
arcmin$^{2}$  (i.e. slightly  larger  than the  detector  field of  view
because  of the  frame dithering).   In practice,  the useful  area is
smaller  since  the  $9\arcmin \times 7\farcm5$  region  around  the
bright star $\rho$  Cen was masked after the  source extraction.  This
corresponds   to  a   density  of   optical  sources   of  $\mu$ = 0.016
arcsec$^{-2}$,  with  $r$ = 4\arcsec--6$\farcs03$.   This  yields  to  a
probability of  chance coincidence between  55 \% and 83 \%, which means
that, at  our limiting  magnitudes, contamination effects cannot be  ignored.
Thus, it is possible  that several of the candidate counterparts  are
indeed spurious  matches.  This conclusion is circumstanced  by Fig.~\ref{pcc},
where  we show the  dependence of the chance coincidence probability $P$
on the position uncertainty.

\subsection{Sources without candidate optical counterparts}

For  32  sources in  our  serendipitous  X--ray  source catalogue  the
cross--matching did not produce any candidate optical/NIR counterpart.
For  seven of  them, \#357,  380,  387, 230,  173, 181,  and 124,  the
apparent lack of matches is ascribed to the fact that they fall within
$\sim$ 6\arcmin\ from the position of  the bright star $\rho$ Cen, i.e.
in a  region which was masked  before running the  source detection on
the  \WFI\ images  (see  \S5.3).   For these  sources  we checked  the
original  unmasked  single--band  optical  catalogues and  we  visually
inspected  the  \WFI\  images  to  verify the  existence  of  possible
counterparts.  For all  of them we found indeed  one or more candidate
optical counterparts  on the \WFI\ images.  However,  since their flux
measurements  are highly uncertain,  they are  useless for  a reliable
X--ray  source identification.   This  is likely  true  also for  flux
measurements taken from, e.g.  the \gsc\ and \tmass\ catalogues, which
were probably affected by the same problem.  Thus, although we spotted
out  their  detection as  a  reference  for  future follow-up  optical
observations, these  candidate counterparts are not  considered in the
following analysis.   The remaining 25 X--ray sources  ($\sim$20\% of
the total) fall  well outside the masked region and  are thus the only
ones which actually lack a candidate optical/NIR counterpart.

\section{X--ray source classification}\label{sec:7}

\subsection{The classification scheme}\label{scheme}

For  all  X--ray  sources  we computed  the  \textit{X--ray--to--optical}  flux
ratio $\frac{f_{X}}{f_{opt}}$. We computed the X--ray flux by assuming
the best--fit emission model and hydrogen column density or, when none
of the  tested models  gives acceptable spectral  fits or  no spectral
fitting  is   possible,  an absorbed power--law   spectrum  with  photon--index
$\Gamma$ = 1.7 and  $N_{H}$ = 1.3$\times$10$^{21}$ cm$^{-2}$, corresponding
to the hydrogen column density  measured in the direction of \1E.  The
optical flux $f_{opt}$ was computed from the measured magnitudes using
the relations reported  in Appendix B of La  Palombara et al.  (2006).
The $\frac{f_{X}}{f_{opt}}$ was mostly computed using the {\em R}--band
magnitude  as  a reference,  because  it was  the  band  with the  most
detections.   When no  {\em R}--band  magnitude was  available  for the
candidate optical counterpart, we  alternatively used the {\em V}, {\em
B}, {\em I},  and {\em U}--band magnitudes (in this  order).  In order to
use the  $\frac{f_{X}}{f_{opt}}$ ratio as a diagnostic  for the X--ray
source   classification,   we   adopted   the   scheme   proposed   by
\citet{LaPalombara2006},       where        sources       with       a
$log(\frac{f_{X}}{f_{opt}}) > 1$  are   likely  extra--galactic,  while
sources with $log(\frac{f_{X}}{f_{opt}}) < -1.5$ are likely stars.  As
a general rule,  in cases where two or  more different spectral models
provide equally acceptable fits to the X--ray spectrum, and thus cause
ambiguity in  the determination of  the $\frac{f_{X}}{f_{opt}}$ ratio,
the  source  classification was  claimed  on  the  basis of  the  best
agreement  between the different  classification indexes  (see below).
When   no  candidate   optical   counterpart  is   found  within   the
cross--matching  radius  we adopted  the  $R$-band limiting  magnitude
($R$ = 23.97)    to     estimate    the    lower     limits    on    the
$\frac{f_{X}}{f_{opt}}$ ratio.

We then used the combined available multi-wavelength information, i.e.
the  best-fitting X--ray  spectra (or  the {\em  HR} for  the faintest
sources),   the  measured   hydrogen  column   density   $N_{H}$,  the
X--ray--to--optical   flux  ratio  $\frac{f_{X}}{f_{opt}}$,   and  the
optical/NIR  colours  of the  candidate  counterparts,  to propose  an
optical identification and a  likely classification for the 112 X--ray
sources selected after the catalogue cross--matching (see \S6.2).  For
the 25  certified sources without candidate  optical counterparts (see
\S6.3) we used the lower limit on the $\frac{f_{X}}{f_{opt}}$ ratio to
support the proposed classifications  based on the source spectrum and
$N_{H}$.  In  some cases,  X--ray source variability  was taken  as an
important classification  index. We  note that, due  to the  quite low
declination of  our field ($\sim -52^\circ$), no  coverage is provided
by available large  scale radio surveys, like the  {\em NVSS} and {\em
FIRST}, and no candidate  radio source counterpart could be identified
which could provide a further classification evidence.

\begin{figure}[!h]
\includegraphics[width=7.5cm,angle=0]{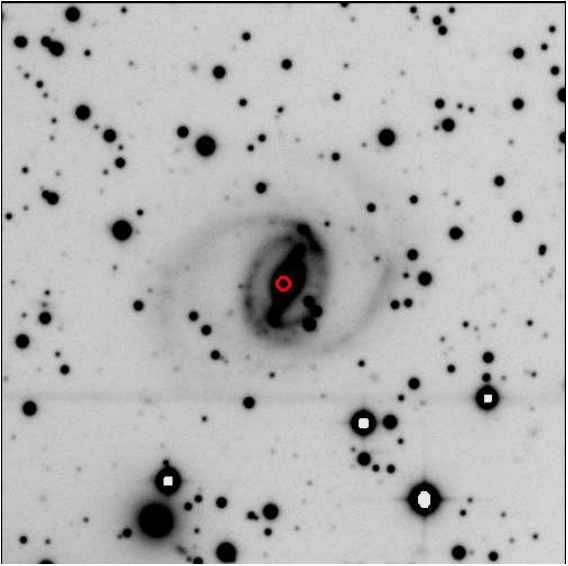}
\caption{$2\arcmin \times 2\arcmin$ {\em R}-band  image of the
Seyfert--2 galaxy ESO  217-G29 taken with the {\em  WFI} at the ESO/MPG
2.2m telescope. The position of the X--ray source \#239 (XMMU J121029.0$-$522148)
is marked with the red circle  (1$\farcs$38 radius) and coincides with the ESO
217-G29 nucleus.}\label{sy2_zoom}
\vspace{-0.5 truecm}
\end{figure}

\subsection{Brightest X-ray sources}\label{sec:bright}

We first  evaluated the  classification of the  X--ray sources  in our
bright  sub sample  (see~\ref{sec:4}),  for  which  the  relatively
accurate determination  of the  source spectrum and  $N_{H}$ represent
already an important piece of  evidence. In addition, for most of them
the optical candidate counterparts are expected to be bright enough to
be detected in  nearly all  the passbands,  and thus  to have  a more
reliable colour-based classification.

As    mentioned    in \S~\ref{sec:sy2},    source    \#239    (XMMU
J121029.0$-$522148) was already  identified in Paper I as  a Seyfert--2
galaxy,  positionally coincident  with  the galaxy  ESO 217-G29.   The
positional coincidence  is further strengthened by  our updated X--ray
coordinates   $\alpha_{J2000}$   =   12$^{h}$  10$^{m}$   29.01$^{s}$,
$\delta_{J2000}$ =  -52$^{\circ}$ 21$'$ 48\farcs1  (after applying the
astrometric correction,  see \S6.1).   The association of  source XMMU
J121029.0$-$522148 with the galaxy ESO 217-G29 is evident in our \WFI\
images  (see Fig.~\ref{sy2_zoom}),  which clearly  resolve  the galaxy
structure (nucleus,  bar, and  spiral arms) and  show that  the source
position  is clearly  coincident with  the bright  nucleus.  Strangely
enough,  the  cross--correlation with  the  \WFI\  catalogue yields  a
candidate optical  counterpart which is at 3\farcs37  from the nominal
X--ray source  position.  This  is an error  of {\em  SExtractor}, the
software used to  run the source detection on  the \WFI\ images, which
did not correctly resolve the nucleus of the galaxy. We thus discarded
the flux of the galaxy computed  by {\em SExtractor} and we assumed an
{\em R}--band magnitude of 14.93,  as reported in {\em
Simbad}. From the computed X--ray  flux (see \S4.3)  we thus
derived      an      X--ray--to--optical      flux      ratio
$\frac{f_{X}}{f_{opt}}=0.166$, in agreement  with the expectations for
a low--luminosity Seyfert--2 galaxy.

In Table~\ref{sources} we listed all the candidate counterparts to the
other 39 X--ray  sources of our bright sample  (see \S~\ref{sec:4}). 32 of
them ($\sim$ 82 \%)  have   at  least  one   optical  candidate
counterpart.  In  particular, for 12 X--ray sources  ($\sim$ 27 \%) the
cross--matching produced more  than one optical candidate counterpart.
For  each of  the  optical candidate  counterparts  (either single  or
multiple) we  reported both their magnitudes (in one  reference passband) and
their $\frac{f_{X}}{f_{opt}}$  ratios (computed for  the assumed X--ray
spectral   model). The proposed classification, reported in
Table~\ref{sources},  is  considered   virtually  secured   when  best
agreement is found between  the different classification indexes, i.e.
the  X--ray source  spectrum  and the hydrogen column density $N_{H}$, on one  side, and  the
colour-based classification  and $\frac{f_{X}}{f_{opt}}$ ratio  of the
optical  candidate  counterpart, on  the  other one.   For simplicity,  we
considered only two main X--ray source classes, i.e. \textit{STELLAR} and
\textit{AGN}:  in the  first class  we include  the  standard galactic
sources  with a soft,  mainly thermal  spectrum and  low X--ray/optical
flux  ratio, while  in  the second  class  we include  extra--galactic
sources   with  a   hard,  likely   non--thermal  spectrum   and  high
X--ray/optical flux ratio. None of our X--ray sources is associated  with cluster of galaxies or with non--active galaxies. 
We flagged cases where the source classification is likely, but not secured, or uncertain because of one
or more inconsistencies  between the different classification indexes.
To this  aim, we  devised the classification
flag {\em a} when the source  classification is likely but not secured by
the identification  of its optical counterpart, since the candidate optical
counterpart is unclassified, or poorly classified, or undetected. Moreover, when
compelling evidence is lacking we consider the source classification as uncertain
with the  following classification flags: {\em  b} when the best--fit
$N_{H}$  value is too low  for AGNs and
too high  for stars; {\em c} when the X--ray spectrum is  not in agreement
either with the \textit{X--ray--to--optical}  flux
ratio or with the  colour-based classification of
the optical candidate counterparts; {\em d} when the source X--ray spectrum
is not unambiguously determined, and/or the  spectral parameters have
large errors.  Of course,  multiple flags were assigned when different
cases apply.

Following  a  decision-tree  approach,  we  thus  proposed  a
virtually secure or likely classification  for 15 of the 39 brightest
X--ray sources  (36 \% of the  total). According to  our classification
scheme, we proposed  that these 15 sources are  active galactic nuclei
(AGNs).  These  sources have  all a clear,  or generally  most likely,
{\em power--law}  X--ray spectrum, relatively  high $N_{H}$, and  6 of
them have an optical candidate  counterpart identified with a QSO, with
a  consistent $\frac{f_{X}}{f_{opt}}$ ratio.
For example, we  classified  source  \#216  (XMMU J120955.1$-$522105)
as  an  AGN, without any flag, because  of  its  {\em  power--law} spectrum  and  $N_{H}$,  and
because  its candidate optical counterparts is classified  as QSO.
We thus  considered the classification  of these  6 sources  as secured.   Three sources,
i.e. \#304 (XMMU J121052.9$-$522354),  \#326 (XMMU J121034.6$-$522457), and
\#517 (XMMU   J121031.9$-$523046),   have   no   optical   candidate
counterpart, while other sources, i.e. \#520  (XMMU J121101.5$-$523030),
\#471  (XMMU J121057.3$-$522905),  and \#533  (XMMU J121013.2-523123),
have  a candidate optical  counterpart but  for which no colour--based
classification  is possible.  However,  their {\em  power--law} X--ray
spectra, $N_{H}$,  and the constraints  on the $\frac{f_{X}}{f_{opt}}$
ratio, suggest  that they are  AGNs.  Furthermore, two of  them, i.e. \#326
(XMMU  J121034.6$-$522457) and  \#520  (XMMU J121101.5$-$523030),  also
feature a significant long term  X--ray variability (see \S4.2), which
reinforces  their  classification as  AGNs.   We thus classified
these five sources as AGNs and we flagged them as {\em a} because of the  lack of  a
possible, or  unambiguous, optical identification.

For 18 X--ray sources  the proposed classifications  reported in
Table~\ref{sources}  (11 AGNs and 7 stars) are  uncertain
because  of inconsistencies between  the classification  indexes.  For
instance, we  classified source  \#158 (XMMU J121018.4$-$521911)  as a
star since its X--ray light curve features long and short flares (see
\S4.2) and its candidate optical  counterpart is an M3 star.  However,
because of  its somewhat large best--fit $N_{H}$, we prudently flagged
its  classification as  {\em  b}. Instead,  source \#198  (XMMU
J120841.6$-$522026)  was classified  as  an AGN  because  of its  {\em
power--law} spectrum,  but it has a  quite low $N_{H}$  and we flagged
its  classification as  {\em b}.   We classified  sources  \#410 (XMMU
J120921.0$-$522700) and \#688 (XMMU J120959.0$-$523618) as AGNs but we
flagged these classifications as {\em a} since their candidate optical
counterparts are unclassified. The former  was also flagged as {\em d}
since its X--ray spectrum is not unambiguously determined.

For the 6 X--ray sources for which no fit to the X--ray
spectrum was possible  with the tested single model  component, or
different  model fits yield comparable $\chi^{2}$
(flagged  with {\em  ``uncl''}  in Table~\ref{sources})  we could  only
suggest,  at  most, tentative  classifications. For instance,  source
\#263 (XMMU J120928.2$-$522225) might  be classified as a galaxy since
the  colours   of  its  nearest  optical   candidate  counterpart  are
consistent with  an elliptical galaxy.  Similarly,  source \#121 (XMMU
J120901.3$-$521741) has a candidate  QSO optical counterpart and might
be   thus   classified   as   an   AGN.   For   source   \#386   (XMMU
J121043.1$-$522638)  not  even the  optical  candidate counterpart  is
classified.   Source  \#357  (XMMU  J121113.8$-$522532)  has  no
candidate counterpart  in our optical/NIR  catalogue\footnote{We note
that this source falls in a region polluted
by the halo of the bright star $\rho$ Cen, which was masked before the
source extraction  (see \S5.3), so that no  match was produced  by the
X--correlation (see \S6.3).  Although a star is indeed detected in the
\WFI\  images, close  the X-ray  source position,  it is  saturated in
almost all bands so that not  even crude optical flux estimates can be
obtained.}.
Source \#426 (XMMU J121000.0$-$522747) remains unclassified, due to 
conflicting power--law spectral model and stellar X-ray/optical flux ratio 
(although within its error-circle a clear galaxy can be seen in the \WFI\ 
images).

Based on the previous analysis, we can summarize the classification of the 39 brightest sources as follows:
\begin{itemize}
\item 15 sources are classified: 5 of them were already classified in Paper I, while 1 had an uncertain classification and 3 were unclassified; the remaining 6 sources are new detections
\item 18 sources have an uncertain classification: 2 of them were classified in Paper I, while 5 were uncertain and 6 unclassified; the remaining 5 sources are new detections
\item 6 source are unclassified: 1 of them was unclassified also in Paper I, while the remaining 5 sources are new detections
\end{itemize}

Our classification analysis improves and supersedes that carried out in Paper I where, apart from the Seyfert--2 galaxy ESO 217-G29, a  classification was proposed only  for 7 of the remaining 23  brightest X--ray sources (30 \%). For these 7 sources we have now revised the  classification  proposed  in Paper I,  which is now confirmed for only 5 of them, while it is downgraded as  uncertain for the other 2. Among the 6 sources with an uncertain classification in Paper 1, one is now firmly classified, while the classification of the other 5 remains uncertain. Finally, 3 of the 10 unclassified sources in Paper I are now classified, while the classification of other 6 is uncertain, and only one still remains unclassified.

We note that the use of the X--ray--to--optical flux ratio, defined in \citet{LaPalombara2006}, as a classification evidence is reliable. For instance, 6 of the 7 sources classified as stars have $log(\frac{f_{X}}{f_{opt}}) < -1.5$, 
while 6 of the proposed AGNs have $log(\frac{f_{X}}{f_{opt}}) > +1$. These values are indeed in agreement with the classifications proposed for X--ray sources detected in the \XMM~\textit{Serendipitous Survey} \citep{Barcons2007}, where most of the identified sources have $-1 < log(\frac{f_{X}}{f_{opt}}) < 1$, and stars and extragalactic sources have the lowest and highest values, respectively.

\subsection{Faintest X--ray sources}

We  also evaluated the  classification of  the 104  remaining, fainter
X--ray sources in  our serendipitous catalogue. Since for  all of them
the lower  number of counts ($\le$ 500) does not allow us to perform an
accurate  spectral  analysis,   the  characterisation  of  the  X--ray
spectrum  only relies on  the source  {\em HR}. As in the case of the bright
sources (\S~\ref{sec:bright}),  the proposed X--ray  source classifications
is based  on the source {\em  HR} and on  the X--ray--to--optical flux
ratio $\frac{f_{X}}{f_{opt}}$, using the  classification scheme
devised  in \citet{LaPalombara2006}.  When  a reliable 
classification of the optical/NIR candidate counterparts was found, we also
used this information as  a further classification evidence.  Based on
the  {\em HR} distribution, we  assumed that
sources with an {\em HR} $<$ -0.9 have spectra corresponding to coronal
emission  from normal stars,  while sources  with {\em HR} $>$ -0.5 are
either  extra--galactic  (normal  or  active galaxies  or  cluster  of
galaxies) or  accreting binary systems  (XRBs or CVs). Because  of the
typical  {\em  HR} errors,  we  considered  sources with  intermediate
values (-0.9 $<$ {\em  HR} $<$ -0.5) as borderline  cases and thus  we did
not   considered   this    parameter   compelling   for   our   source
classification.  For sources affected by  too large errors on the {\em
HR} this parameter was not considered at all. As in \S~\ref{scheme},
when  no candidate counterparts  were found  we assumed  the $R$ = 23.97
limiting   magnitude   of  the   \WFI~catalogue   to  compute   the
$\frac{f_{X}}{f_{opt}}$ lower limit.

Following  the  same  decision-tree  approach  used  to  classify  the
brightest  X--ray sources,  4  of  the 25  sources  with no  candidate
counterpart remained unclassified, while  all the remaining 21 sources
were identified with  an AGN. 
 
On the other hand, among the 36 sources with a single candidate 
counterpart 10 were identified as stars (2 sure and 8 uncertain), 19 as 
AGNs (8 sure and 11 uncertain) and 2 with galaxies (since the \WFI\ images 
show an evident extended source as countepart). The other 5 sources 
remained unclassified, due to unconstrained or conflicting hardness ratio 
and/or X--ray/optical flux ratio, but in the error--circle of two of them a 
clear galaxy can be seen in the \WFI\ images. Finally, in the  case of
the  43 X--ray  sources  with  two or  more  candidate counterparts  we
proposed 11  classifications as stars  (10 sure and only  1 uncertain)
and 28 classifications  as AGNs (27 sure and  only 1 uncertain), while
for the other 4 sources we were unable to suggest any classification.

To summarize,  we classified 21 sources  (corresponding to 20 \% of the
total) as stars and 68 sources (65 \%) as AGNs, while other 2 sources (2 \%) were identified with galaxies and the remaining 13 sources (13 \%) remained unclassified. We note that 17 of the sources classified as stars have $log(\frac{f_{X}}{f_{opt}}) < -1.5$.
On the other hand, 16 sources classified as AGNs have a high X--ray-to-optical flux ratio $log(\frac{f_{X}}{f_{opt}}) > +1$.
As in the case of the bright sources (see~\ref{sec:bright}), our X--ray-to-optical flux ratios yield classifications which are in agreement  with those similarly proposed for X--ray sources detected in other surveys \citep{Barcons2007}.

\section{Summary    and    conclusions}

We  analysed all the  \XMM~observations of  the intermediate--latitude
field around 1E1207.4$-$5209 in order to investigate the properties of
the X--ray source population. We detected 144 serendipitous sources in
total; 114 of them were detected in the soft energy band (0.5--2 keV),
while 87 were detected in the  hard energy band (2--10 keV) band, down
to limiting fluxes of $\sim$10$^{-15}$ erg cm$^{-2}$ sec$^{-1}$ and 4$\times$10$^{-15}$ erg cm$^{-2}$ sec$^{-1}$, respectively.   The lower
number of  fainter sources detected  with respect to that  reported in
Paper I  (see \S  2.4) mainly affects  the log$N$--log$S$ distribution in  the soft
energy band, which now features a clear flattening at the low flux end
(i.e.   below $\sim$ 4$\times$10$^{-15}$ erg cm$^{-2}$ sec$^{-1}$).
However, at higher fluxes the log$N$--log$S$ distribution is perfectly
consistent  with that  reported in  Paper I  and is  well  above those
obtained at  high galactic latitudes  \citep{Baldi2002}.  We therefore
confirm  the   presence  of  a  non   negligible  galactic  population
component, in addition to the extra--galactic one.  In the hard energy
band, the  log$N$--log$S$ distribution  is fully consistent  with that
reported in Paper I and with those obtained both in the Galactic plane
\citep{Ebisawa2005} and  at high Galactic  latitude \citep{Baldi2002},
confirming  that  the  distribution  is dominated  by  extra--galactic
sources.  Thanks to the increased count statistics, we could perform a
variability and spectral analysis of  the 40 brightest sources. For 10
of them, we  spotted a large flux variation between  the 2002 and 2005
observations, suggesting  that they  are transient sources,  while for
other two we found evidence of variability on short timescales ($\sim$
0.1 and $\sim$ 10 ks).   Moreover, we refined the spectral analysis of
the Seyfert--2 galaxy XMMU J121029.0$-$522148 we discussed in Paper I,
finding a best--fit  redshift value $z$ = 0.042,  higher than the value of 0.032
reported in the literature. We also carried out a complete multi--band
({\em UBVRI}) optical coverage  of the  field with  the \WFI\  of the
ESO/MPG 2.2m telescope to search for candidate optical counterparts to
the  X--ray sources  and we  found  at least  a candidate  counterpart
brighter than $V \sim 24.5$ for 112 of them.  By cross--identification
with sources in the  \tmass~catalogue, we also provided a colour--based
classification  for  most of  them.   We  thus  identified 27  of  the
brightest sources  as AGNs and 7  as stars, while we  identified 21 of
the  faintest  sources  as  stars  and 70  sources  as  AGNs or galaxies.   Future
follow--up   works   will  be   aimed   at   confirming  the   proposed
classification  of the  brightest X-ray  sources  through multi--object
spectroscopy of  the candidate counterparts. For the  proposed AGNs we
also  plan   to  perform  radio  observations  to   achieve  a  better
classification.

\section*{Acknowledgments}
\begin{acknowledgements}
This work is based on  observations obtained with \XMM, an ESA science
mission  with instruments  and  contributions directly  funded by  ESA
Member States  and NASA.  The  \XMM~data analysis is supported  by the
Italian  Space  Agency  (ASI).  This  publication makes  use  of  data
products from the Two Micron All  Sky Survey, which is a joint project
of  the University of  Massachusetts and  the Infrared  Processing and
Analysis  Center/California  Institute of  Technology,  funded by  the
National Aeronautics and Space Administration and the National Science
Foundation. RPM acnowledges STFC for support through a Rolling Grant
and INAF - IASF Milano for hospitality.

\end{acknowledgements}

\bibliographystyle{aa}
\bibliography{biblio}

\begin{sidewaystable*}
\caption{Main characteristics of the 39 brightest sources. The sources are sorted by decreasing count number.}\label{sources}
\begin{tabular}{cc|ccccc|ccccc|c} \hline \hline
(1)	& (2)		& (3)	& (4)		& (5)			& (6)			& (7)			& (8)		& (9)	&(10)		& (11)	    &(12)		& (13)		\\
SRC	& NAME  & cts	& Model		& N$_{\rm H}$		& $\Gamma$/kT		& $\chi_{\nu}^{2}$	& D$_{XO}$	& MAG 	& OPTICAL		& $\frac{f_{X}}{f_{opt}}$	  & OPTICAL     & X-RAY SOURCE		\\
	&  &	&		& (10$^{21}$ cm$^{-2}$)	& (-/keV)		&			& (arcsec)	& (mag) &	FILTER	& (log$_{10}$)    &  COUNTERPART CLASS		&	CLASS	\\ \hline
520	& XMMUJ121101.5-523030	& 6464	& wabs(pow)	& 2.2$^{+0.4}_{-0.3}$ & 1.97$^{+0.12}_{-0.11}$ & 1.07 & 2.71 & 19.68 & R & 0.39 & - & AGN$^{a}$		\\ 
338	& XMMUJ120942.1-522458	& 5611	& wabs(pow)	& 1.1$^{+0.3}_{-0.3}$ & 1.98$^{+0.12}_{-0.11}$ & 1.15 & 2.99 & 19.43 & R & 0.15 & QSO & AGN		\\ 
241	& XMMUJ120858.8-522129	& 3829	& wabs(mekal)	& 1.4$^{+0.6}_{-0.6}$ & 0.62$^{+0.02}_{-0.02}$ & 1.63 & 3.07 & 13.71 & V & -2.11 & - & STAR$^{a,c}$ \\ 
508 & XMMUJ120857.1-523014  & 2287  & wabs(brem)	& 2.9$^{+0.9}_{-0.8}$ & 0.28$^{+0.70}_{-0.54}$ & 1.81 & 3.14 & 16.25 & R & -1.55 & - & STAR$^{a,c}$ \\ 
    &                       &       & wabs(bbody)	& 1.0$^{+0.9}_{-0.6}$ & 0.17$^{+0.02}_{-0.02}$ & 1.72	&  &  &   &  &  & 		\\ 
480	& XMMUJ120908.1-522918  & 2146	& wabs(pow)	& 1.2$^{+0.5}_{-0.5}$ & 1.92$^{+0.19}_{-0.16}$ & 0.90 & 2.42 & 19.36 & R & -0.06 & QSO & AGN	\\ 
198	& XMMUJ120841.6-522026	& 1976	& wabs(pow)	& 0.1$^{+0.02}_{-0.04}$ & 1.85$^{+0.22}_{-0.19}$ & 0.95 & 3.38 & 20.14 & R & 0.46 & QSO & AGN$^{b}$	\\ 
    &                       &       &           &                     &                        &      & 3.28 & 21.71 & R & 1.10	& MS M3-M4 & 	\\ 
674	& XMMUJ121007.2-523555	& 1631	& wabs(pow)	& 0.8$^{+0.5}_{-0.5}$ & 1.71$^{+0.14}_{-0.17}$ & 1.13 & 1.83 & 19.26 & R & -0.39 & QSO & AGN$^{b}$	\\ 
244 & XMMUJ120842.5-522128  & 1603  & wabs(pow) & $<$0.08 & 1.66$^{+0.20}_{-0.22}$ & 0.82 & 3.70 & 21.97 & V & 1.22 & - & AGN$^{a,b}$		\\ 
    &                       &     	&           &         &                        &      & 3.33 & 21.02 & R & 0.75 & MS A7-WD & 		\\ 
357	& XMMUJ121113.8-522532	& 1488	& wabs(brem)	& 4.4$^{+1.8}_{-1.5}$ & 0.26$^{+0.10}_{-0.10}$ & 1.73 & - & - & - & - & - & uncl$^{c}$		\\
	& 		                & 	    & wabs(bbody)	& 2.5$^{+1.9}_{-1.4}$ & 0.16$^{+0.03}_{-0.03}$ & 1.75 & - & - & - & - & - & (see note 7)	\\ 
471 & XMMUJ121057.3-522905  & 1353  & wabs(pow)	& 9.4$^{+0.4}_{-2.8}$ & 2.12$^{+0.32}_{-0.33}$ & 1.48 & 3.92 & 20.11 & R & -0.10  & - & AGN$^{a}$		\\ 
    &                       &       & wabs(mekal)	& 8.9$^{+3.4}_{-2.8}$ & 3.63$^{+2.06}_{-1.06}$ & 1.42 & 3.60 & 20.42 & R & 0.03 & - & 		\\ 
    &                       &   	& wabs(brem)	& 7.6$^{+2.8}_{-2.0}$ & 4.27$^{+3.28}_{-1.53}$ & 1.48 &  &   &  &  & - & 		\\ 
509	& XMMUJ120913.8-523023	& 1319	& wabs(pow)	& 0.9$^{+0.5}_{-0.5}$ & 1.97$^{+0.17}_{-0.21}$ & 0.89 & 2.67 & 20.11 & R & 0.00 & WD & AGN$^{b,c}$		\\ 
    &                       &   	& wabs(brem)	& 0.3$^{+0.4}_{-0.0}$	& 4.27$^{+2.57}_{-1.24}$ & 0.96 &	 &  &   &   &         & 		\\ 
426	& XMMUJ121000.0-522747	& 1271	& wabs(pow)	& 0.1$^{+0.6}_{-0.0}$	& 1.64$^{+0.16}_{-0.18}$ & 2.15	& 3.66 & 15.84 & R & -1.92 & Sbc-vB2 & uncl$^{c}$		\\ 
340	& XMMUJ120951.1-522525	& 1131	& wabs(pow)	& 0.4$^{+0.7}_{-0.0}$	& 1.79$^{+0.21}_{-0.28}$ & 0.87	& - & $>$23.97	& R & $>$1.24 & - & AGN$^{a,b}$		\\ 
    &                       &   	& wabs(brem) & $<$0.4	& 5.34$^{+4.37}_{-2.10}$ & 0.93	& - &   &   &  & - & 		\\ 
\hline \hline
\end{tabular}
\vfill
\end{sidewaystable*}

\addtocounter{table}{-1}
\begin{sidewaystable*}
\caption{Continued}\label{}
\begin{tabular}{cc|ccccc|ccccc|c} \hline \hline
(1)	& (2)		& (3)	& (4)		& (5)			& (6)			& (7)			& (8)		& (9)	&(10)		& (11)	    &(12)		& (13)		\\
SRC	& NAME  & cts	& Model		& N$_{\rm H}$		& $\Gamma$/kT		& $\chi_{\nu}^{2}$	& D$_{XO}$	& MAG 	& OPTICAL		& $\frac{f_{X}}{f_{opt}}$	  & OPTICAL     & X-RAY SOURCE		\\
	&  &	&		& (10$^{21}$ cm$^{-2}$)	& (-/keV)		&			& (arcsec)	& (mag) &	FILTER	& (log$_{10}$)    &  COUNTERPART CLASS		&	CLASS	\\ \hline
598	& XMMUJ120927.4-523326	& 1128	& wabs(pow)	& 0.9$^{+0.6}_{-0.5}$	& 2.07$^{+0.16}_{-0.21}$ & 0.92	& 3.81 & 20.17 & V & 0.00 & - & AGN$^{a,b}$	\\ 
404	& XMMUJ121017.5-522706	& 1119	& wabs(mekal)	& 5.3$^{+1.1}_{-1.2}$	& 0.63$^{+0.06}_{-0.06}$ & 1.39	& 3.46 & 15.47 & R & -2.51 & MS K3-K2 & STAR$^{b}$	\\ 
158	& XMMUJ121018.4-521911	& 1090	& wabs(brem)	& 2.8$^{+1.5}_{-1.2}$	& 0.28$^{+0.11}_{-0.08}$ & 1.38	& 4.04 & 16.49 & R & -1.82 & MS M3 & STAR$^{b}$	\\
490 & XMMUJ120935.6-522940  & 1058 	& wabs(pow)	& 4.5$^{+2.6}_{-1.5}$	& 2.46$^{+0.64}_{-0.45}$ & 1.09	& 3.00 & 22.33 & R & 0.33 & QSO & AGN		\\ 
    &                       &       & wabs(brem)	& 3.0$^{+1.9}_{-1.3}$	& 2.46$^{+1.93}_{-0.84}$ & 1.10	&  &  &  &  &	-	& 		\\ 
    &                       & 	    & wabs(bbody)	& 0.2$^{+1.9}_{-0.0}$	& 0.57$^{+0.09}_{-0.09}$ & 1.29	&  &  &  &  &	-	& 		\\ 
216	& XMMUJ120955.1-522105	& 1033  & wabs(pow)	& 1.5$^{+0.8}_{-0.9}$	& 2.06$^{+0.37}_{-0.29}$ & 1.25	& 3.39 & 18.74 & R & -0.78 & QSO & AGN	\\
    &                       &       &           &                       &                        &      & 3.58 & 22.27 & R & 0.63 & MS M2 & 	\\
585	& XMMUJ120906.9-523310	& 980	& wabs(pow)	& 1.8$^{+0.8}_{-0.8}$	& 1.96$^{+0.21}_{-0.25}$ & 1.01	& 3.79 & 21.76 & R & 0.58 & QSO & AGN		\\
    &		                &	    & wabs(brem) & 1.0$^{+0.8}_{-0.5}$	& 4.46$^{+3.58}_{-1.48}$ & 1.05	& 2.53 & 19.51 & R & -0.32 & QSO & 		\\ 
141	& XMMUJ120945.2-521828	& 959	& wabs(pow)	& 0.1$^{+0.5}_{-0.0}$	& 1.48$^{+0.28}_{-0.18}$ & 1.38	& 3.46 & 21.05 & R & 0.44 & QSO & AGN$^{b}$	\\ 
688	& XMMUJ120959.0-523618	& 946	& wabs(pow)	& 2.3$^{+1.0}_{-0.9}$	& 1.94$^{+0.22}_{-0.29}$ & 0.80	& 2.18 & 22.00 & V & 0.57 & - & AGN$^{a}$		\\ 
    &                       &   	& wabs(brem) & 1.5$^{+1.0}_{-0.6}$	& 4.69$^{+4.41}_{-1.58}$ & 0.81	& 3.63 & 22.53 & V & 0.79 & - & 		\\ 
654	& XMMUJ120936.5-523515	& 888	& wabs(pow)	& 1.3$^{+0.1}_{-0.0}$	& 2.08$^{+0.69}_{-0.55}$ & 1.63	& 3.38 & 22.66 & R & 0.89 & - & AGN$^{a}$		\\
    &	                    &	    &           &                    	&                        &   	& 1.51 & 22.88 & R & 0.98 & - & 		\\
433	& XMMUJ120850.5-522738	& 818	& wabs(brem) & 2.2$^{+0.2}_{-0.1}$	& 0.47$^{+0.21}_{-0.17}$ & 1.69	& 2.27 & 16.63 & R & -1.77 & MS K5-M0 & STAR$^{b}$		\\
214	& XMMUJ121040.9-522055	& 785	& wabs(pow)	& $<$0.3	& 1.92$^{+0.24}_{-0.21}$ & 1.76	& -	& $>$23.97	& R & $>$1.39 &  - & AGN$^{a,b}$		\\ 
363	& XMMUJ121039.9-522538	& 773	& wabs(mekal) & 1.7$^{+1.1}_{-1.4}$	& 0.53$^{+0.10}_{-0.06}$ & 2.70 & 3.06 & 14.22 & R & -2.74 & - &  STAR$^{a,b}$		\\ 
304	& XMMUJ121052.9-522354	& 699	& wabs(pow)	& 2.4$^{+2.6}_{-1.4}$	& 1.56$^{+0.38}_{-0.28}$ & 1.41	& - & $>$23.97	& R & $>$1.52 & - & AGN$^{a}$		\\ 
326	& XMMUJ121034.6-522457	& 694	& wabs(pow)	& 7.8$^{+5.0}_{-3.7}$	& 1.30$^{+0.30}_{-0.31}$ & 1.31	& - & $>$23.97	& R & $>$1.44 & - & AGN$^{a}$		\\ 
    &                       &   	& wabs(bbody)	& 0.9$^{+0.3}_{-0.0}$ & 1.26$^{+0.24}_{-0.18}$ & 1.33 & - & $>$23.97 & R & $>$1.40 & - & 		\\     
372 & XMMUJ120950.5-522613  & 618 & wabs(bbody) & $<$0.05 & 0.25$^{+0.03}_{-0.03}$  & 1.26 & 5.43 & 20.57 & R & -0.55 & MS K4 & STAR$^{d}$   \\   
    &                       &     & wabs(brem) & $<$0.05 & 0.73$^{+0.27}_{-0.24}$ & 1.13 &2.59 & 23.86 & U & 0.82 & - &                      \\
386	& XMMUJ121043.1-522638	& 605	&     ?	        &     -             &            -             &   -  & 2.41 & 18.47 & R & -0.96 & - & uncl		\\
236	& XMMUJ120904.8-522129	& 551	& wabs(pow)	 & 2.2$^{+1.7}_{-1.6}$	& 1.89$^{+0.66}_{-0.28}$ & 0.94	& 3.50 & 22.42 & R & 0.65 & - & AGN$^{a}$		\\ 
    &                       &       & wabs(bbody) & $<$0.9	& 0.61$^{+0.13}_{-0.11}$ & 1.21	&   &   &   &   &  & 		\\ 
\hline \hline
\end{tabular}
\vfill
\end{sidewaystable*}

\addtocounter{table}{-1}
\begin{sidewaystable*}
\begin{minipage}[t][180mm]{\textwidth}
\caption{(Continued) }\label{} 
\begin{tabular}{cc|ccccc|ccccc|c} \hline \hline
(1)	& (2)		& (3)	& (4)		& (5)			& (6)			& (7)			& (8)		& (9)			& (10)	    &(11)		& (12)	& (13)	\\
SRC	& NAME  & cts	& Model		& N$_{\rm H}$		& $\Gamma$/kT		& $\chi_{\nu}^{2}$	& D$_{XO}$	& MAG & OPTICAL	& $\frac{f_{X}}{f_{opt}}$	  & OPTICAL       & X-RAY SOURCE		\\
	&  &	&		& (10$^{21}$ cm$^{-2}$)	& (-/keV)		&			& (arcsec)	& (mag) & FILTER	& (log$_{10}$)    &  COUNTERPART CLASS		&	CLASS	\\ \hline
222 & XMMUJ121006.3-522122  & 540	& wabs(pow)	& 0.5$^{+1.5}_{-0.0}$	& 2.38$^{+0.93}_{-0.51}$ & 1.06	& 3.36 & 20.44 & R & -0.50	& QSO & AGN$^{b}$		\\ 
    &                       &       & wabs(brem) & $<$0.7	& 1.70$^{+1.30}_{-0.73}$ & 1.04	& 2.18 & 19.76 & R & -0.77 & QSO &              		\\ 
204 & XMMUJ120934.1-522034  & 535	& wabs(brem) & 6.4$^{+5.6}_{-3.1}$	& 4.35$^{+8.50}_{-2.53}$ & 1.56	& 3.48 & 21.39 & R & 0.21 & MS F5 & uncl$^{c}$		\\ 
    &                       &       & wabs(pow)	& 8.9$^{+0.8}_{-4.6}$	& 2.18$^{+1.02}_{-0.37}$ & 1.53	&  &  &  & 0.23 & - & 		\\ 
    &                       &       & wabs(bbody)	& 0.7$^{+4.7}_{-0.0}$	& 0.77$^{+0.18}_{-0.19}$ & 1.72	&  &  &  & 0.14 & - & 		\\ 
410 & XMMUJ120921.0-522700  & 535	& wabs(pow)	& 4.6$^{+0.5}_{-0.3}$	& 2.15$^{+0.58}_{-0.69}$ & 1.54	& - & $>$23.97 & R & $>$0.88 & - & AGN$^{a,d}$		\\ 
    &                       &	    & wabs(bbody)	& 0.2$^{+0.3}_{-0.0}$ & 0.65$^{+0.18}_{-0.18}$ & 1.75 & - &  &  & & - & 		\\ 
121 & XMMUJ120901.3-521741  & 533	&     ?  	&         -             &            -            &   -   & 3.75 & 21.87 & R &0.49 & QSO & uncl		\\ 
106 & XMMUJ120955.3-521716  & 522	& wabs(pow) & 4.8$^{+0.4}_{-2.4}$   & 1.73$^{+0.57}_{-0.41}$ & 1.16 & 2.95 & 19.96 & R & -0.18 & QSO &  AGN		\\ 
    &                       &       &           &                       &                        &      & 2.13 & 23.53 & R & 1.24 & - & 		\\ 
263 & XMMUJ120928.2-522225  & 515	&     ?	   &          -             &            -           &  -   & 0.86 & 22.46 & R & 0.49 & - & uncl		\\ 
    &                       &       &     ?	   &                        &                        &      & 2.19 & 22.51 & R & 0.51 & QSO & 		\\ 
    &                       &	    &     ?    &                       &                       &   	& 3.93 & 22.09 & R & 0.34 & MS M2.5 & 	\\ 
    &                       &	    &     ?    &                       &                       &     & 3.04 & 22.14 & R & 0.36 & MS K7-K5 & 	\\ 
64  & XMMUJ120958.7-521449  & 512	& wabs(pow)	& 0.2$^{+0.1}_{-0.2}$  & 1.84$^{+0.79}_{-0.46}$ & 1.25 & 1.72 & 23.46 & R & 1.26 & - & AGN$^{a,b}$		\\ 
517 & XMMUJ121031.9-523046  & 508	& wabs(pow)	 & 1.7$^{+0.4}_{-0.0}$	& 1.78$^{+0.86}_{-0.58}$ & 0.77	& - & $>$23.97 & R & $>$0.73 & - & AGN$^{a}$		\\ 
    &                       &	    & wabs(bbody)	& $<$1.6	& 0.63$^{+0.18}_{-0.14}$ & 0.98	& - &  &  &	 &  & 		\\ 
533 & XMMUJ121013.2-523123  & 499	& wabs(pow)	 & 3.4$^{+0.3}_{-0.2}$	& 2.05$^{+0.30}_{-0.44}$ & 1.43	& 3.65 & 20.66 & U & -0.18 & - & AGN$^{a}$		\\ 
    &                       &	    & wabs(bbody)	& $<$1.4 & 0.65$^{+0.11}_{-0.10}$ & 1.63 & 3.28 & 16.18 & R & -2.25 & - &             	\\ \hline \hline
\end{tabular}
\vfill

\begin{tiny}
Key to  Table -  Col.(1): source ID  number.  Col.(2)  catalogue name.
Col.(3):  source  total  counts  (in  the 0.3--8  keV  energy  range).
Col.(4): best--fit emission model(s); the symbol `uncl' indicates that
none  of the  tested single--component  models provided  an acceptable
fit.  Col.(5):  best--fit hydrogen column density  with the associated
90 \%  confidence errors.  Col.(6): best--fit  photon--index or plasma
temperature   (for  a   power--law  or   a  thermal   emission  model,
respectively)  with the  associated  relevant 90\% confidence  level
error.   Col.(7):  best--fit  reduced chi--square.   Col.(8):  angular
distance between the X--ray  source position and its optical candidate
counterpart (if  any). The most likely optical  counterpart are listed
first.   Col(9): magnitude  of  the optical  candidate counterpart  or
$R \ge$ 23.97   upper    limit   if   no    candidate   counterpart   is
found. Col.(10): optical filter;  if the optical candidate counterpart
has  no  $R$--band magnitude  we  considered the  V,  B,  I and  U--band
magnitudes  in  this  order.   Col.(11):  logarithmic  values  of  the
X--ray--to--optical  flux ratio;  the  optical flux  is  based on  the
magnitudes in Col.(9) while the X--ray flux is based on the best--fit
model or, when  no model is acceptable, on  a power--law spectrum with
photon--index  $\Gamma$  = 1.7  and  hydrogen  column density  N$_{\rm
H}=1.3\times 10^{21}$  cm$^{-2}$, corresponding to  the total galactic
column  density.  Col.(12):  suggested classification  of  the optical
candidate  counterpart from the  {\em WFI}  catalogue. MS indicates a
main sequence star, QSO indicates a quasar, WD indicates a white dwarf, 
Sbc and vB2 indicate a spiral galaxy and a blue compact galaxy respectively.
Col.(13):  proposed source  classification  of
X--ray source with warning flags: {\em a} the source classification is
likely  but   not  secured  by  the  identification   of  its  optical
counterpart:  the candidate  optical counterpart  is  unclassified, or
poorly classified, or undetected; {\em b} the source classification is
uncertain because the best--fit $N_{H}$  value is too low for AGNs and
too high  for stars, {\em c}  the X--ray spectrum is  not in agreement
either with  the magnitude or with the  colour-based classification of
the candidate optical counterparts, {\em d} the source X--ray spectrum
is not  unambigously determined,  and/or the spectral  parameters have
large errors.
\end{tiny}

\end{minipage}
\end{sidewaystable*}

\end{document}